\documentclass[twocolumn,aps,prx]{revtex4-1}
\usepackage{ifpdf}
\usepackage{times}
\pdfoutput=1
\usepackage{amsmath}
\usepackage{amssymb}
\usepackage{units}
\usepackage[T1]{fontenc}
\usepackage[applemac]{inputenc} 	
\usepackage{graphicx}
\usepackage{dcolumn}
\usepackage{bm}
\usepackage{color}

\def\be{\begin{equation}}
\def\ee{\end{equation}}
\def\bea{\begin{eqnarray}}
\def\eea{\end{eqnarray}}

\newcommand{\YBCO}{YBa$_2$Cu$_3$O$_{x}$}

\renewcommand{\vec}[1]{\boldsymbol{#1}}

\begin{document}
\preprint{0}

\title{Protected superconductivity at the boundaries of charge-density-wave domains}
\author{Brigitte Leridon}
\email[Corresponding Author: ]{Brigitte.Leridon@espci.fr}
\affiliation{LPEM, ESPCI Paris, CNRS, Universit\'e PSL, Sorbonne Universit\'es, 10 rue Vauquelin, 75005 Paris, France}
\author{Sergio Caprara}
\affiliation{Department of Physics, Sapienza University of Rome, Piazzale Aldo Moro 2, I-00185, Rome, Italy}
\author{J. Vanacken}
\author{V.V. Moshchalkov}
\affiliation{KULeuven, Celestijnenlaan 200 D, B-3001 Leuven, Belgium}
\author{Baptiste Vignolle}
\affiliation{LNCMI (CNRS, EMFL, INSA, UJF, UPS), Toulouse 31400, France}
\email[Present address: ]{CNRS, Univ. Bordeaux, ICMCB, UMR5026, F-33600 Pessac, France}
\author{Rajni Porwal}
\author{R.C. Budhani}
\email[Present address: ]{Morgan State University, 119 Calloway Hall, 1700E Cold spring lane, Baltimore MD 21251, USA}
\affiliation{CSIR-National Physical Laboratory, New Delhi-110012, India}
\author{Alessandro Attanasi}
\author{Marco Grilli}
\author{Jos\'{e} Lorenzana}
\email[Corresponding author: ]{jose.lorenzana@cnr.it}
\affiliation{ISC-CNR and Department of Physics, Sapienza University of Rome, Piazzale Aldo Moro 2, I-00185, Rome, Italy}

%

\begin{abstract}
\textbf{
Solid $^4$He may acquire superfluid characteristics due to the frustration of the solid phase at grain boundaries. 
Here, we show that an analogous effect occurs in systems with competition among charge-density-waves (CDWs) and 
superconductivity in the presence of disorder, as cuprate or dichalcogenide superconductors. The CDWs breaks apart 
in domains with topologically protected filamentary superconductivity (FSC) at the interfaces. 
Transport experiments carried out in underdoped cuprates with the magnetic field acting as a control parameter are shown to be in excellent agreement with the theoretical expectation. 
At high temperature and low fields we find a transition from CDWs to 
fluctuating superconductivity, weakly affected by disorder, while at high field and low temperature the protected filamentary superconducting phase appears  
in close analogy with ``glassy'' supersolid phenomena in $^4$He.
}
\end{abstract}

\maketitle
Electrons in the presence of attractive interactions crossover smoothly from the Bardeen-Cooper-Schrieffer (BCS) 
limit to the Bose condensation (BC) limit as the strength of the interaction is increased \cite{Nozieres1985}.  
However, as electrons approach the limit of composite bosons, the tendency to localize in real space also increases. 
Thus, in analogy with $^4$He, a real-space ordered state competes with a momentum-space condensed state. 
Since the entropy of these states is equally small, phase stability is insensitive to temperature, resulting in 
a phase boundary nearly parallel to the $T$ axis and perpendicular to any blue non-thermal control 
parameter axis (pressure, strain, magnetic field, doping, etc.). 

The scenario changes dramatically in the presence of real-space disorder. It has been known for some time that 
a polycrystal of $^4$He atoms develops superfluidity at the interface and acquires supersolid characteristics 
(i.e., superfluid-like changes of the moment of inertia coexisting with real-space order). See Ref.\,\cite{Balibar2008} 
for a review. It is natural to expect that the analogous phenomenon should occur for real-space fermion 
pairs \cite{Attanasi2008}. In this work, we consider a simple phenomenological model which allows to study the effect 
of quenched disorder near a transition from a real-space ordered state of fermion pairs to a superconducting (SC) 
state. We show that disorder induces filamentary superconductivity (FSC) in the spatially ordered charge-density-wave 
(CDW) state analogous to the supersolid behavior in $^4$He. A finite temperature phase diagram is derived. This 
theoretical scenario is explored experimentally by transport experiments in La$_{2-x}$Sr$_x$CuO$_{4}$ (LSCO) 
using magnetic field as a tuning parameter for dopings close to the insulator-SC transition. 
We show that the experimentally derived transition lines are in agreement with the theoretical expectations:
At moderate temperatures there is a magnetic field driven transition between 
the superconductor and the CDW phase, as seen with other probes \cite{Gerber2015}.
Lowering the temperature in the CDW phase, a very fragile SC phase
appears  due to the coherent phase-locking of SC filaments at the interfaces of CDW domains analogous to the 
supersolid effects seen in $^4$He.

\begin{figure}[tb]
\includegraphics[width=0.9\linewidth, angle=00, clip]{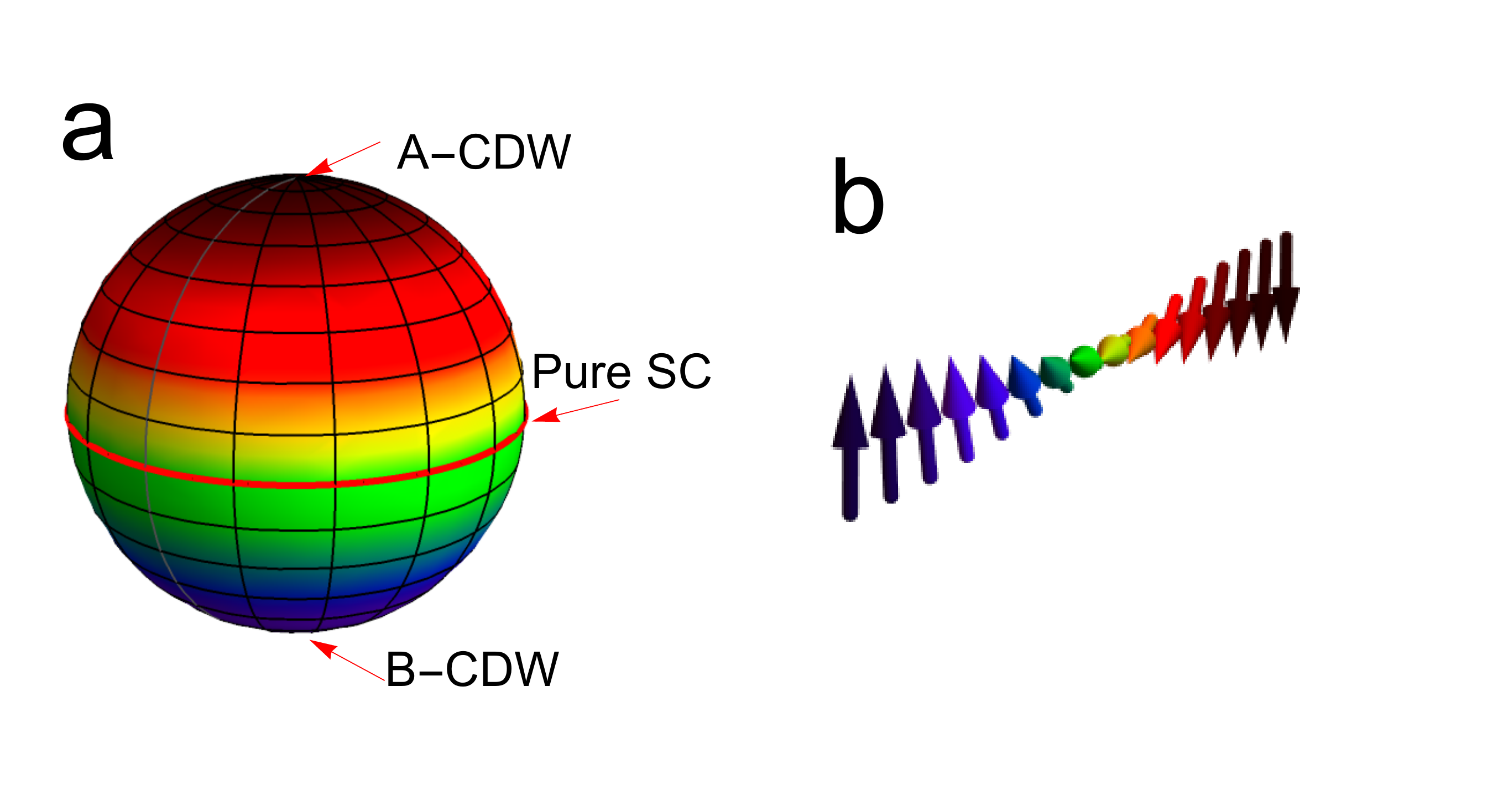} 
\caption{{Pseudospin representation for the possible states.} (a) The sphere represents the states encoded by 
the order parameter with near SO(3) symmetry. North and south poles represent the two possible CDW states, 
corresponding to the charge maximum in one of the two possible sublattices, while the equator encodes the SC state. 
The azimuthal angle encodes the phase of the order parameter. (b) Schematic pseudospin pattern at an interface 
between B/A-CDWs.}
\label{fig:sphere}
\end{figure}

\section*{Results}

\subsection*{Theory of disorder-induced filamentary superconductivity on charge-density-waves}

We consider an electronic system with an attractive interaction that favors real-space formation of fermion pairs. 
At low temperatures these pairs can either condense in a SC state or form a CDW. An instructive example to study 
this competition (or eventual intertwining \cite{Fradkin2015}) is the negative-$U$ 
Hubbard model in a bipartite lattice. At half-filling the CDW and superconductivity become degenerate. The model 
supports two variants of the CDW (labeled A and B) differing on which of the two sublattices hosts more charge than 
the other. 

At strong coupling, the negative-$U$ Hubbard model can be mapped \cite{Micnas1990} onto a Heisenberg model with 
pseudospins describing charge degrees of freedom analogous to Anderson's pseudospins \cite{Anderson1958}, but in 
real space instead of momentum space. Assuming an ordered state at $T=0$, the projection of the staggered 
magnetisation on the \textit{xy}-plane encodes the superfluid order parameter, while its $z$ component encodes the 
CDW order parameter. A positive (negative) component describes the A-CDW (B-CDW) (Fig.\,\ref{fig:sphere}). In the 
Hubbard model, the order parameter has SO(3) symmetry, which reflects perfect degeneracy between CDW 
and superconductivity. This degeneracy is not generic and  gets broken when other interactions beyond the canonical 
Hubbard model are considered. For example, a nearest-neighbor repulsion (attraction) favors the CDW (SC) and drives 
the effective model to have Ising (\textit{xy}) symmetry \cite{Micnas1990}. 

It is natural to assume that in a region of parameter space in which charge order and superconductivity are seen 
to coexist, this near degeneracy is reestablished and a model with near SO(3) symmetry generically describes a 
situation in which the energy for real-space (i.e. CDW) or momentum-space (i.e. SC) condensation of paired fermions is similar. 
A related lattice model has been considered by Liu and Fisher to study possible supersolid phases in 
$^4$He \cite{Liu1972} close to the boundary between the crystalline phase and the superfluid phase underling the 
close analogy with our problem. 

Since we are interested in intertwining, and in the finite temperature phase diagram, we will neglect 
quantum fluctuations. These become important when the temperature is below the characteristic energies 
of the problem, which, especially in the FSC region, are very low. Thus we will study a semiclassical model of 
CDW-SC in the spirit of Ref.\,\cite{Liu1972} for the supersolid problem.

\begin{figure}[tb]
\includegraphics[width=8cm, angle=0, clip]{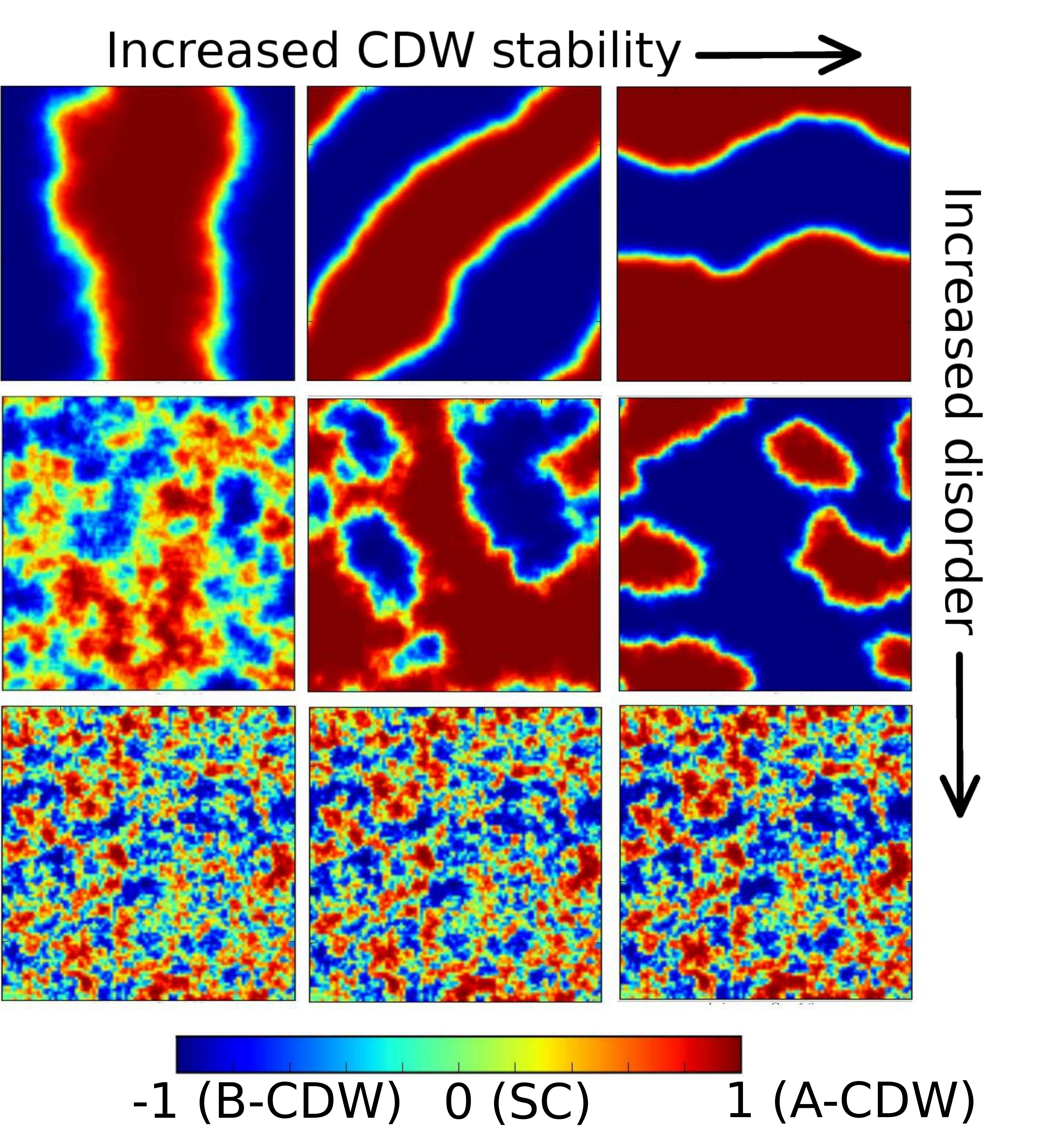}
\caption{
Zero temperature configurations obtained by minimizing the energy functional Eq.\,(\ref{eq:ham}). 
The control parameter is taken to be $G>0$, so in the absence of disorder the system is in the CDW phase. The 
size of the system is  $100 \times100$. The false color plots represent $S_z$; blue and red regions correspond to 
the two CDWs and light green regions are the SC regions.  
For every row the control parameter $G$ increases from left to right
$G/J = 0.02, 0.06, 0.1$. For every column disorder increases  
from top to bottom with the following strength:
$W/J = 0.5, 1.5, 4.0$.}
\label{fig:config}
\end{figure}

\begin{figure*}[tb]
\includegraphics[width=15cm,  angle=00, clip]{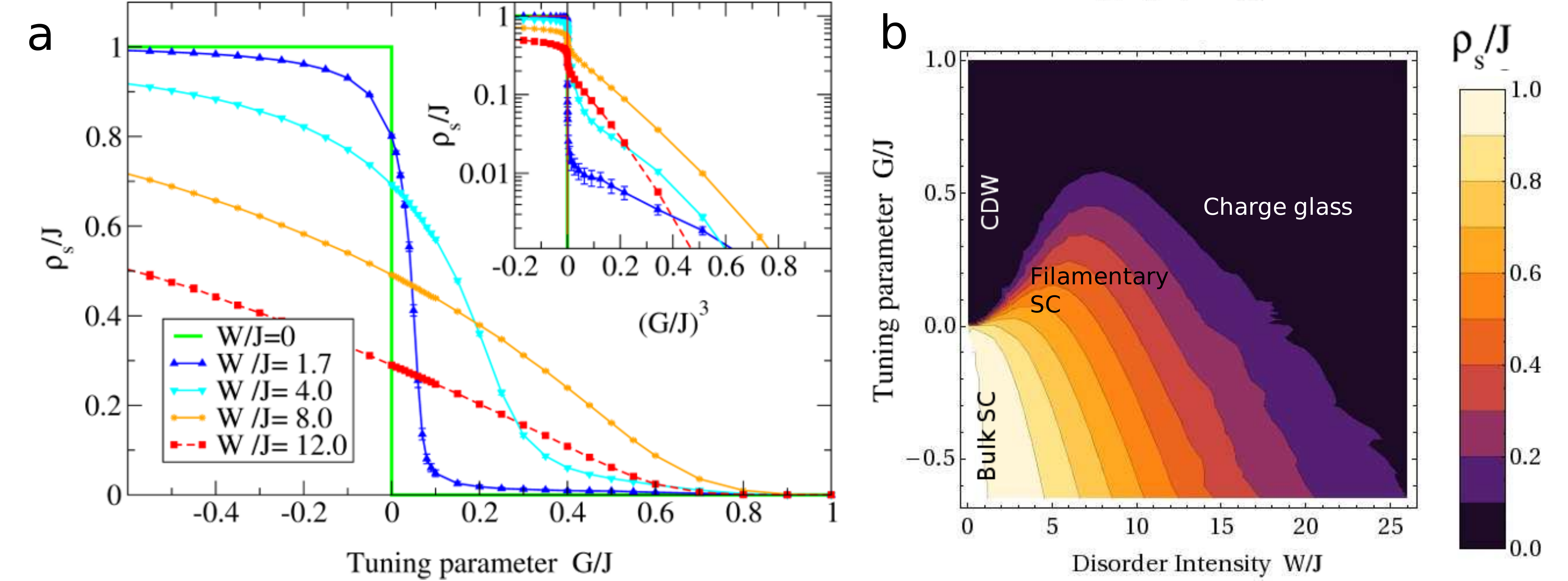}
\caption{(a) Zero temperature stiffness as a function of the tuning parameter $G/J$
for different strengths of disorder and system size $100 \times100$ sites as in Fig.\,\ref{fig:config}.
The inset shows the same quantity in semi-log scale as a function of $(G/J)^3$.
(b) False color plot of the stiffness as a function of tuning parameter and disorder.    
}
\label{fig:phasestif}
\end{figure*}

A generic semiclassical model can be justified by a coarse grained process. 
Assuming that there is at least short range order in the system we can separate it in regions larger than the 
lattice spacing but smaller than the correlation length and define a coarse grained ordering field $\vec{S}_i$ 
which determines the kind of order in region $i$. We neglect the fluctuations in the strength of fermion pairing 
which is parametrised by the length of the ordering field so we take $|\vec{S}_i|=1$. Thus we assume that the CDW 
consists of localized bosons which in the context of insulator-SC transitions is associated with a Mott-Hubbard 
bosonic insulating phase\cite{Cha1991}. We also neglect all complications due to unconventional symmetry of the 
order parameter. In addition, we assume that there are only two possible variants of CDW phases as for the Hubbard 
model so $S^z= 1$  ($S^z= -1$)  encodes the A(B)-CDW. In the Discussion section below, we enumerate possible 
microscopic origins of the different CDW variants in different materials. Our considerations are, however, independent of these 
microscopic details. A pure SC state is described by the complex ordering field $S^x_i+iS^y_i$ with $S^z= 0$, 
while sideways configurations describe the CDW analog of supersolid behavior as in Ref.\,\cite{Liu1972}.
Fig.\,\ref{fig:sphere}(a) displays the order-parameter space. We define the semiclassical model on a discrete lattice 
of cells which is convenient for numerical simulations,
\begin{equation}
\label{eq:ham}
H=-J\sum_{<i,j>}\vec{S}_i\cdot\vec{S}_j-G\sum_i(S_i^z)^2+\sum_ih_i S_i^z.
\end{equation}
Here, $J>0$ describes a short-range stiffness which, for simplicity, we choose to be equivalent for SC correlations 
and CDW correlations.  In the absence of disorder ($h_i=0$), the balance between the orders is decided by the 
parameter $G$. In this case $G>0$ describes a uniform CDW while $G<0$ describes a uniform superconductor. 
One could as well have used an anisotropic Heisenberg model with the same scope, which at the classical level 
we are considering would only change minor details. $h_i$ is a random variable that takes into account that charged 
impurities
will locally favour the A- or B-CDW, depending on whether the impurities on the cell $i$ have more charge near the A or the B sublattice. 
 We will take the $h_i$ to be random variables 
with a flat probability distribution between $-W$ and $W$ and, since we are interested in layered systems 
(cuprates, dichalcogenides), we will consider a two-dimensional (2D) system.

For $G>0$  the model falls into the universality class of the random-field Ising model. As such, for any disorder, 
it breaks apart in domains of the A- and B-CDW variants. This is obvious for large disorder while for small 
disorder it follows from  Binder's refinement \cite{Binder1983} of Imry's and Ma's arguments \cite{Imry1975}. In 
the latter case, however, domains can be exponentially large [roughly $\propto \exp(J^2/W^2)$].

For $G>0$ one can consider a flat interface between an A-CDW and a B-CDW. Since the only way for the spin to reverse 
is by passing through the equator, the interface is forced to have the ordering field on the $xy$ plane and is 
locally SC as shown schematically in Fig.\,\ref{fig:sphere}(b). 
By minimizing the energy, one finds that the SC region has width $\xi_g=\xi_0\sqrt{J/G}$ 
where $\xi_0$ is a short-range cutoff of the order of the coarse grained lattice spacing corresponding to the 
correlation length of the short-range SC (i.e., particle-particle) or CDW (i.e., particle-hole) pairs.
Thus, although for $G>0$  the superconductor is globally less stable, it gets stabilized locally because of 
topological constraints. In the interface both CDW are frustrated so the less stable SC phase prevails as in 
polycrystalline  $^4$He.

Adding disorder to the model for $G>0$ makes the uniform CDW to break apart in domains as expected from Imry's, Ma's 
and Binder's arguments \cite{Imry1975,Binder1983}. Figure\,\ref{fig:config} shows configurations obtained by 
minimizing the functional Eq.\,(\ref{eq:ham}) at $T=0$, with different disorder strengths (see Methods).  Blue and red corresponds 
to the A- and B-CDW respectively. For small $G>0$ and small disorder (upper left corner), one has large domains. 
As expected, FSC nucleates at the interfaces (light green). From left to right, the local CDW stability (controlled 
by $G$) increases. FSC regions become narrower as $\xi_g\sim 1/\sqrt{G}$. Scrolling down the figure, disorder 
increases, producing a decrease in the size of the CDW domains, which favours the formation of a dense FSC 
network. The SC phase is uniform along the FSC regions so that at $T=0$, if the interfaces form a percolative path, 
the system has zero resistivity. 

In Fig.\,\ref{fig:phasestif}(a) we show the zero temperature superfluid stiffness as a function of the tuning 
parameter $G$ for different disorder strengths. We define the stiffness from the second derivative of the energy 
with respect to a twist in the boundary conditions and compute it using linear response theory and performing 
a configuration average \cite{Attanasi2008}. For zero disorder (light green) the stiffness jumps from the bare value 
to zero as the systems changes abruptly from the SC state to the CDW at $G^*=0$. However, as disorder becomes finite 
(blue curve) the stiffness develops a ``foot''  for positive $G$ indicating that FSC is induced in the nominally CDW 
region (see Fig.\,\ref{fig:config}). For large detuning from $G^*=0$ phase stiffness is suppressed 
exponentially, roughly as $\rho_s\sim\exp[-C (G/J)^3]$, with $C$ a constant depending on the disorder strength
(see inset). This indicates that an ever more fragile SC regime sets in as the tendency to CDW is increased and 
the filaments forming the network become narrower. We anticipate that the decreasing but finite stiffness will 
produce a characteristic ``foot'' in the temperature dependent phase diagram, which we take as the fingerprint of 
FSC.

\begin{figure*}[tb]
\includegraphics[width=15.0cm, angle=00, clip]{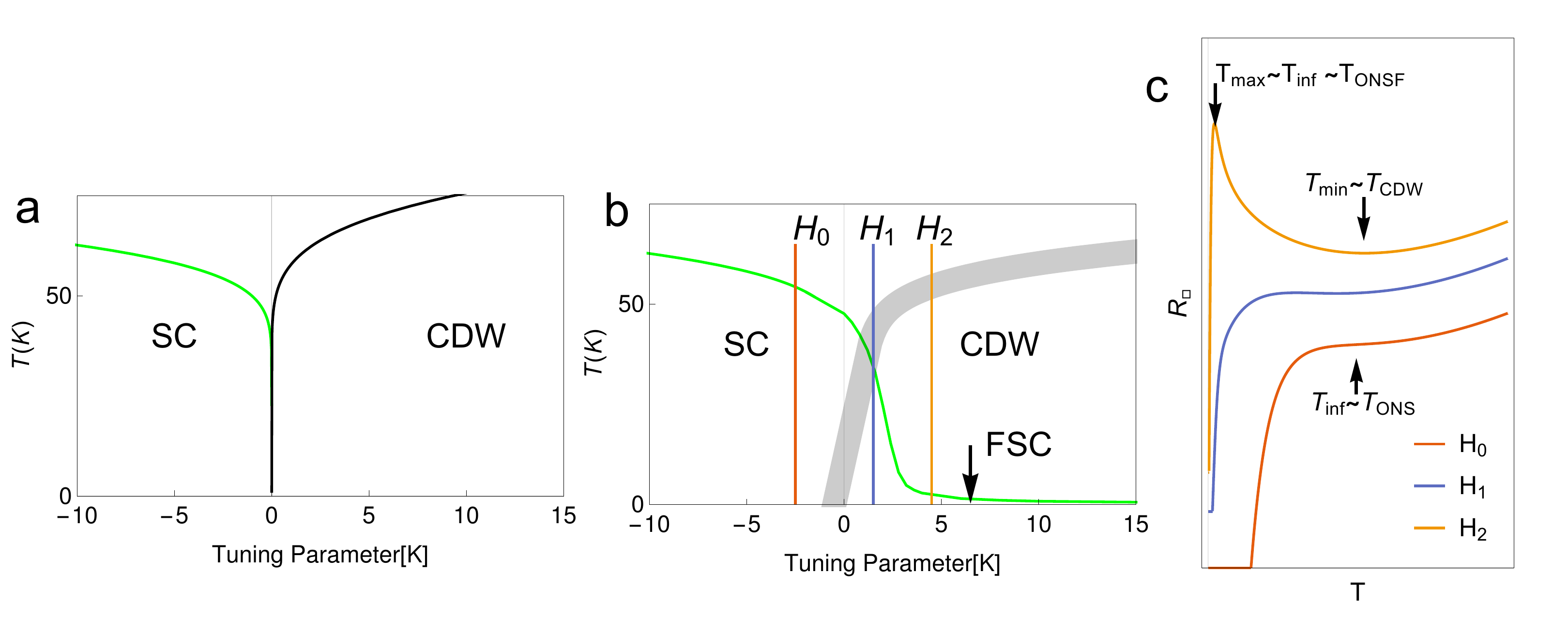}
\caption{Schematic phase diagram in the clean limit (a) and with small disorder (b). The sharp transition line 
to the CDW phase in the clean case (black full line) becomes a crossover in the disordered case (gray band). 
To establish a qualitative connection with transport experiments in LSCO we schematically indicate the role of 
magnetic field as the tuning parameter. In panel (c) we then report the expected behavior of the resistivity 
in the different regions of the (dirty) phase diagram. Curves have been shifted vertically for clarity. 
For $H=H_0$, the resistivity is expected to be a decreasing function of temperature. For $H=H_1$ and high 
temperatures the resistivity will exhibit a plateau due to the residual influence of the clean quantum-critical 
point. Lowering the temperature, disorder becomes relevant and eventually the system becomes SC. For $H=H_2$, 
the resistivity will exhibit a minimum when the CDW correlations set in and a maximum before dropping due to the 
FSC phase. The arrows indicate the characteristic temperatures discussed in the text.}
\label{fig:tcdg}
\end{figure*}

Moderately stronger disorder makes the superfluid phase more robust because the network of filaments becomes denser. 
At some point, however, for very strong disorder, charge localization is favored at every site and the system becomes 
an insulating charge glass. The different regimes can be seen by plotting isolines of the phase stiffness as shown 
in Fig.\,\ref{fig:phasestif}(b). We see that, with increasing disorder, FSC is a reentrant phase in the CDW 
region. Interestingly, for a fixed control parameter $G>0$, there is an optimum value for disorder to 
induce superconductivity. 

Although this phase diagram is for zero temperature, we can derive a finite temperature phase diagram assuming that 
in the clean limit the system behaves as an anisotropic Heisenberg model \cite{Cuccoli2003a} [Fig.\,\ref{fig:tcdg} (a)] 
and in the presence of disorder develops a finite temperature SC phase with a $T_c$ proportional to the $XY$ phase 
stiffness. Fig.\,\ref{fig:tcdg}(b) shows schematically the modified phase diagram in the presence of small disorder. 
Here, we have assumed that $T_c$ is proportional to the stiffness of the $W/J=1.7$ case in the filamentary region 
and interpolates smoothly to the $T_c$ of the clean limit for negative tuning parameter. 
We have chosen the microscopic parameters so that the energy scale represents typical values of underdoped 
cuprates. In the FSC phase the CDW domains coexists with superconductivity.  
In this mixed state, like the ones shown in  Fig.\,\ref{fig:config}, the  CDW transition gets broadened by the effect 
of disorder so that the sharp Ising-like transition of the clean case becomes a crossover (gray band) with 
glassy characteristics in the presence of disorder \cite{Miao2017}. In the case of unidirectional CDW a sharp 
transition may persist in a nematic channel, as discussed in Refs.\,\cite{Nie2014,Capati2015}.   

In order to associate the phase diagram to magnetotransport experiments, we will define below characteristic 
temperatures from resistivity data and assume they can be used as proxies of the different transition or crossover lines. 
A guide to the various temperature scales introduced in this work can be found in Table\,\ref{tabT}.

Since magnetic field $H$ is known to tip the balance between superconductivity and CDW \cite{Gerber2015} we will 
use $H-H^*$  as the tuning parameter where $H^*$ corresponds to the field of a clean quantum-critical point (CQCP), 
i.e., zero tuning parameter ($G^*=0$). Thus, in Fig.\,\ref{fig:phasestif}(a) we associate the abscissa axis with the 
magnetic field (increasing from left to right). In the figure, the temperature units of the tuning parameter 
were approximately derived from equivalent energy units in the microscopic Heisenberg model. Conversion to 
magnetic-field units would require a precise mapping of the models which is beyond our scope. Empirically, 
we find that, as an order of magnitude, 5\,K of the microscopic model corresponds to 10\,T of the experiment.\\

\begin{table}[h]
\begin{tabular}{c|c|c}
Temp.  & Proxy& Meaning  \\
scale & ~ \\ \hline 
~&~\\
$T_c^B$&$T_c$ & $T_c$ for bulk superconductivity  \\
$T_c^F$&$T_c$ & $T_c$ for filamentary superconductivity  \\
$T_{ONS}$& $T_{inf}$& Temp. for the onset of robust SC correlations  \\
$T_{ONSF}$&$T_{inf}$ & Temp. for the onset of FSC effects  \\
$T_{CDW}$& $T_{min}$& CDW crossover temperature  \\
\vspace{-0.1 cm}
$T_0$ &Fit &Cutoff temp. above which an exponential  \\
~ & &suppression of paraconductivity occurs \\
\vspace{-0.1 cm}
$T_1$ &Fit & Characteristic temp. scale for the width of  \\
~ & & low-$T$ suppression of resistance due to FSC \\
~&~\\
\hline
~&~\\
$T_c$ & &Temp. at which $R_\square(T)$ vanishes  \\
$T_{max}$& & Temp. at which a local maximum in $R_\square(T)$ occurs    \\
$T_{min}$& & Temp. at which a local minimum in $R_\square(T)$ occurs   \\
$T_{inf}$& & Temp. at which an inflection point in $R_\square(T)$ occurs 
\end{tabular}
\caption{Guide to the meaning of the various temperature scales introduced in the text.
The horizontal line separates temperature scales that are introduced in the theory and/or the
fitting formulas from temperature scales that are defined by the experimental temperature dependence
of the measured resistance $R_\square(T)$.}
\label{tabT}
\end{table}

At zero or low field [$H_0$, red line in Fig. \ref{fig:tcdg} (b)] the metallic phase directly becomes SC so transport experiments are 
expected to yield a monotonic decreasing function of temperature, as shown with the red curve in panel (c).
We will use the inflection point in this curve $T_{inf}$ [indicated by the arrow in panel (c)] as a function 
of field as a proxy for $T_{ONS}$, the characteristic temperature below which robust in-plane SC correlations 
appear. Notice that three-dimensional (3D) zero-resistance superconductivity sets in at a lower temperature, $T_c$. 
$T_{inf}$ should not be confused with another inflection point appearing around 270\,K at this doping and 
unrelated to superconductivity \cite{Ando2004,Pelc2017}. At intermediate fields, $H=H_1$, and at 
high/intermediate temperatures, when disorder is not yet relevant, the system critically fluctuates between the 
SC and CDW states, giving rise to a flat resistance that would seemingly extrapolate to a zero-temperature
CQCP (located nearly zero tuning parameter). Eventually, however, SC prevails at low $T$ and 
the resistance vanishes (blue lines). At higher fields ($H=H_2$, orange lines) the metal first enters a region 
of disordered (polycrystalline) CDW with an insulating behavior, thus the temperature corresponding to the 
resistivity minimum serves as a proxy of the CDW crossover temperature $T_{CDW}$, as shown in panel (c). Lowering 
even more the temperature the resistivity shows an inflection point very close to a sharp maximum followed, by a 
rapid drop when coherence establishes between the SC filaments. We will use the inflection point (near the maximum) 
to signal the onset of FSC, $T_{ONSF}$, as shown with the arrow in the upper left corner of panel (c). For weak disorder, the 
intermediate region showing clean quantum-critical behavior (i.e., the resistivity plateau) is around the region 
in parameter space where $T_{max}$, $T_{min}$ and $T_{inf}$ merge. 

\subsection*{Filamentary superconductivity in LSCO as revealed by magneto-transport}

\begin{figure*}[tb]
\includegraphics[width=17cm, angle=00, clip]{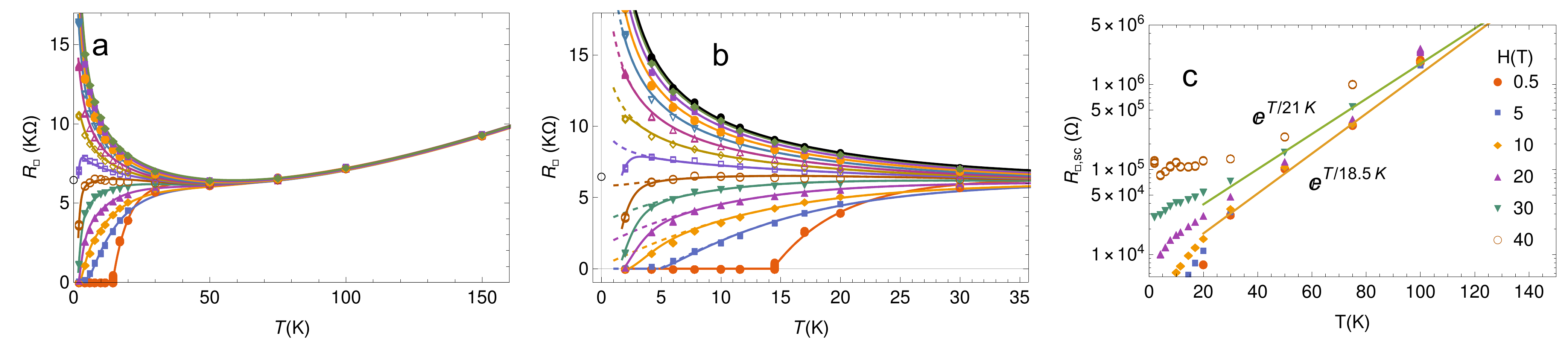}
\caption{Temperature dependent resistivity for sample 008 (LSCO/STO, $x=0.08$) for different 
fields. Panel (a) shows data at selected fields together with the 
2D fits. Fields start from $H=0$\,T in the lower curve up to 48\,T in intervals of 4\,T. Panel (b) 
shows detailed data at low temperatures. The black curve is the fit of the data without SC component at 50\,T 
(black dots). Dashed lines are the same fits but without the $\tanh$ low temperature cutoff (see text).   
In (a) and (b) the open black circle at $T=0$ indicates the quantum of resistance, 
$R_Q=h/(2e)^2$. 
Panel (c) is the superconductivity-related resistance in semi-log scale showing the exponential behavior of 
the resistivity above a cutoff temperature $T_0$, as shown by the lines.} 
\label{008rho}
\end{figure*}

We expect the scenario presented in the previous section to be realized in underdoped cuprates, where 
superconductivity and CDW are known to compete. In particular, when a strong magnetic field \cite{Gerber2015} or 
strain \cite{Kim2018} are present, a static charge order is well documented in the underdoped region of the 
phase diagram, below a temperature $T_{CDW}$  \cite{Kivelson2003,Wu2011}. This charge-ordered phase is believed 
to be responsible for the Fermi surface reconstruction at doping values below $p=0.16$ \cite{Doiron-Leyraud2007}. 
We thus chose to study the resistivity of LSCO thin films with Sr doping slightly above the minimal doping 
for superconductivity (typically $x=0.08-0.09$), in order to be able to drive gradually the system towards the 
insulating state by increasing the magnetic field.  The resistivity of the thin films was measured at 
different temperatures under pulsed magnetic field as high as 54\,T. The Method section reports on the experimental 
details of the transport measurements.

Our goal here is to establish a link between the theoretical analysis and transport experiments carried out in 
the presence of intense magnetic fields. A microscopic computation of transport would be a formidable task as 
it would require to take into account the quantum nature of quasiparticles, their scattering in the CDW 
regions, their role in mediating phase coherence between the SC filaments. In addition one should consider 
the crossover between this low-temperature coherent SC state and the more 2D regime at higher temperature 
in which Cooper pairs fluctuate as in standard 2D metals (giving rise to the observed regime of rather 
robust Aslamazov-Larkin (AL) paraconductivity fluctuations 
\cite{Aslamazov1968a,Aslamasov1968,LeridonPRB2007,Caprara2009}) and quantum fluctuations at low temperatures
\cite{Cha1991}. Therefore, we avoid in our approach any microscopic attempt to describe the resistivity experiments. 
Instead, we find that a relatively simple and physically motivated expression fits the 
resistivity in the whole temperature and magnetic field range as shown in Fig.\,\ref{008rho}(a) and (b) 
and Fig.\,S1 (full lines) and we choose to make use of it to extract relevant characteristic 
temperatures in a phenomenological and systematic way.

The fits were obtained considering two independent contributions to the conductivity, 
\begin{equation}
\label{eq:rdt}
\left[R_\square(H,T)\right]^{-1}= \left[R_{\square,{\rm SC}}(H,T)\right]^{-1}+
\left[R_{\square,{\rm CO}}(T)\right]^{-1} .
\end{equation}
The second term on the r.h.s. represents the field independent conductivity in the absence of any SC fluctuation and is 
characterised by a crossover from the linear high-temperature behaviour of resistivity of the metallic state to 
the logarithmic insulating-like behaviour taking place at low $T$ under strong magnetic fields, that we associate 
to the formation of the polycrystalline CDW state (charge-ordered state). We therefore assume that this is the 
high-field behaviour of the system and estimate it by a fit to the higher field data. The resulting 
$R_{\square,{\rm CO}}(T)$ is shown with a black line in  Fig.\,\ref{008rho}(b). Subtracting this contribution to 
the total conductivity allows to identify the contribution of superconductivity (static as well as 
fluctuating) to transport, as represented by the first term in the right hand side of Eq.\,(\ref{eq:rdt}), which 
also encodes all the significant magnetic field dependence of transport.

The superconductivity-related resistance shows that at high-temperature SC fluctuations disappear rapidly 
(exponentially with reduced temperature) above a characteristic temperature $T_0$, as shown in 
Fig.\,\ref{008rho}(c). This rapid suppression of fluctuations was  previously observed in \YBCO\ 
\cite{LeridonPRL2001, LeridonPRL2003} and LSCO thin films \cite{LeridonPRB2007} and associated to the presence 
of an energy cutoff \cite{Vidal2002,Mishonov2003,Caprara2005} whose value is found to be doping-dependent 
\cite{Luo2003}. This observation has been reproduced recently in various cuprates \cite{popcevic2018}.

At low temperatures and low/intermediate fields the superconductivity-related resistance is expected to 
display a linear behavior in temperature as follows from 2D AL fluctuations 
at temperatures close to $T_c$. We adopt a simple phenomenological form which interpolates between the 
exponential behavior at high temperature and the AL behavior, 
\begin{equation}
\label{eq:rsupb-app}
R_{\square,{\rm SC}}^{AL}(H,T)=\frac{e^{T/T_0}-e^{T_c^B/T_0}}{\sigma_{\square,0}}
\theta(T-T_c^B).
\end{equation}
Indeed this expression behaves as $\sim T-T_c^{B}$ at low temperature where  $T_c^{B}$ will be termed the 
``bulk'' critical temperature. The parameters $T_0$, $T_c^{B}$ and $\sigma_{\square,0}$
are taken to be functions of $H$ as explained in Methods.

The fit with this expression is shown with dashed lines in Fig.\,\ref{008rho}(b). It represents very well the 
experimental data except for the low temperature region ($\lesssim 8$K) at intermediate fields. The root of the 
problem becomes clear upon inspection of panel (a). Notice that fluctuating superconductivity produce a visible 
magnetoresistance below the zero field from $T_{ONS}\approx 55$\,K  all the way down to $T_c$. Thus, the AL regime 
is associated with a very broad regime of fluctuations characteristic of a 2D superconductor.
In contrast, as it is clear from Fig.\,\ref{008rho}(b) and Supplementary Information Fig.\,S1, a much more rapid 
variation  sets in below $8$\,K where points at intermediate fields are not fitted by the dashed lines 
[Eq.\,(\ref{eq:rsupb-app})]. In other words, it is not possible to fit with a single AL form both the broad 
fluctuating regime starting at $T_{ONS}$ and the low-$T$ regime at intermediate fields. This calls for a different 
mechanism setting in at low $T$, which we attribute to FSC short circuiting an otherwise finite low temperature 
resistivity.  In order to describe this effect we simply add an hyperbolic tangent cutoff to the SC component, 
\begin{equation}
\label{eq:rsupffil-app}
R_{\square,{\rm SC}}(H,T)=R_{\square,{\rm SC}}^{AL}(H,T)
\tanh\left(\frac{T-T_c^{F}}{T_1}\right) \theta(T-T_c^{F}).
\end{equation}
The theta in Eqs.\,(\ref{eq:rsupb-app}),\,(\ref{eq:rsupffil-app})
ensures that zero resistivity occurs at the maximum among the ``bulk'' and the filamentary critical temperature 
parameters, $T_c^B$ and $T_c^F$ respectively. Using Eq.\,(\ref{eq:rsupffil-app}) we obtain the full line fits of 
Fig.\,\ref{008rho}, which now work over the whole temperature range. 

\begin{figure*}[htb]
\includegraphics[width=5.5cm,  clip]{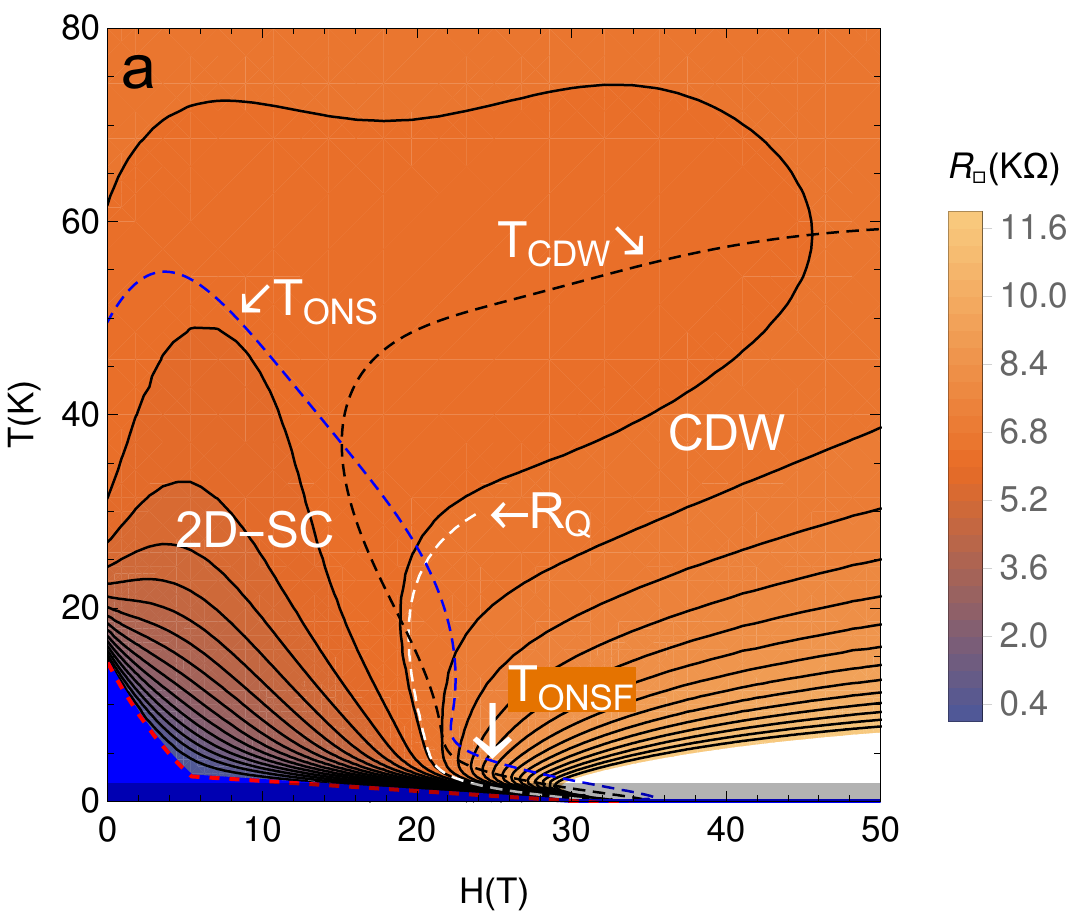}
\includegraphics[width=5.5cm,  clip]{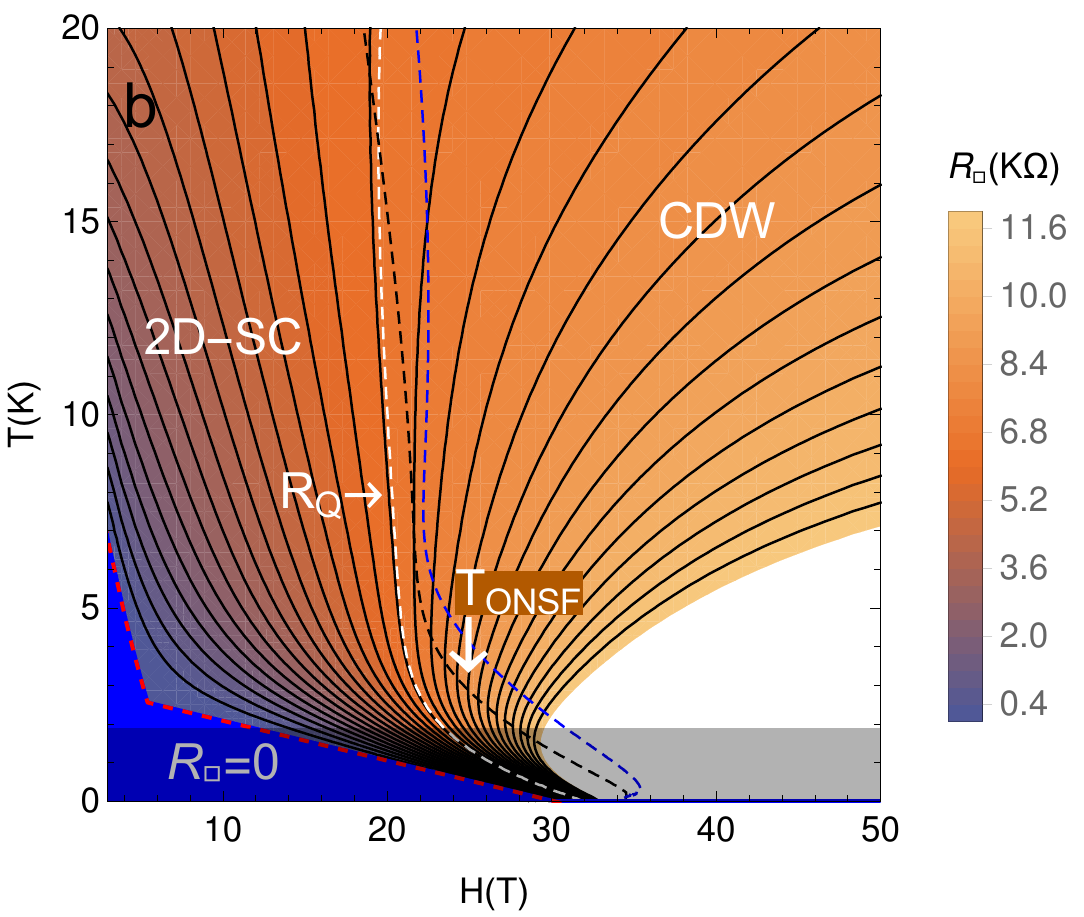}
\includegraphics[width=5.5cm,  clip]{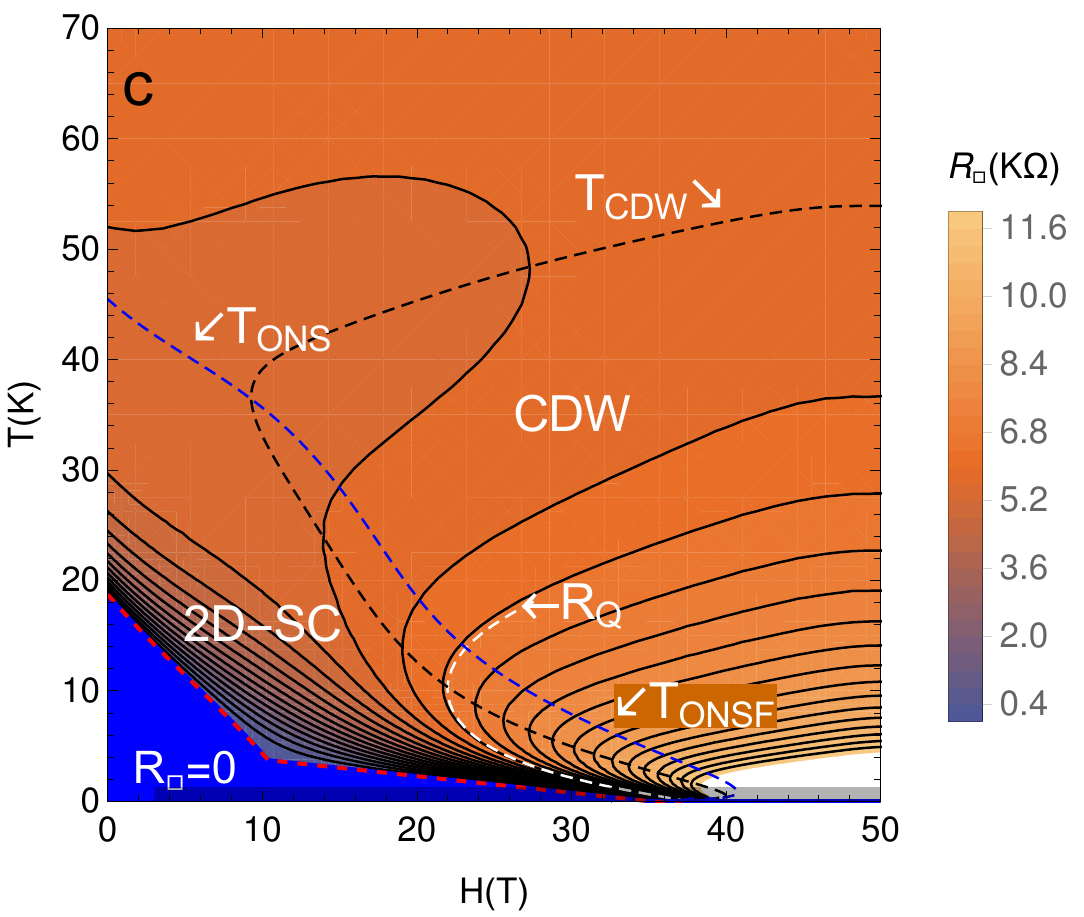}
\caption{Phase diagram from experimental resistivity encoded on a color scale for : a) and b) sample 008 
LSCO/STO, $x=0.08$; c)  sample 009  LSCO/LSAO, $x=0.09$.
The red dashed line is the extrapolated $T_c$ below which $R_\square=0$ (blue region). The black dashed line 
represents the  zeroes of the first temperature derivative of the square resistance with the upper (lower) 
branch representing a minimum (maximum). The blue dashed line represents the zeroes of the second derivative, 
i.e., the inflection point which lays between the minimum and the maximum of each $R_\square(T)$ curve when they 
exists. Full lines are equal resistivity levels in intervals of 0.4\,k$\Omega$ with the lower visible level close 
to $T_c$ corresponding to 0.4\,k$\Omega$. The white dashed line is the isoline corresponding to the quantum of 
resistance. False colors encode the square resistance. The grey rectangle at the bottom indicates the extrapolated 
region (i.e. out of the range of available experimental data). Notice, however, that the leading behavior defining 
the foot is already quite clear from the available data in Fig.\,\ref{008rho}. (b) is a zoom of the critical region 
of (a).  
}
\label{phdiag008}
\end{figure*}

Once the optimum surface is obtained in the $(H,T)$ plane (see Methods), it is possible to {perform equal-resistivity plots} as shown in 
Fig.\,\ref{phdiag008}. Here we also show the maximum between $T_c^B(H)$ and $T_c^F(H)$ (red dashed line) which 
determines the upper boundary of the $R_{\square}(H,T)=0 $ region (colored blue). As expected $T_c^B(H)$ [$T_c^F(H)$] 
is dominant at low [high] field with a noticeable kink at the intersection. In order to {extract} the 
characteristic crossover temperatures defined in Fig.\,\ref{fig:tcdg}(c)  we now determine the temperatures of 
the resistivity maximum $T_{max}(H)$ and minimum $T_{min}(H)$  by solving 
$\partial R_{\square}(H,T)/\partial T=0$. Furthermore, we find the temperature $T_{inf}(H)$ of the resistivity 
inflection point, $\partial^2 R_{\square}(H,T)/\partial T^2=0$ . 

We first discuss the inflection point at low fields (blue dashed line). At high temperature, the resistivity has a  
positive curvature [Fig.\,\ref{fig:phasestif}(a)] which at small field is compensated by the onset of SC
correlations. Therefore the inflection point is taken as the characteristic onset temperature, $T_{inf}\sim T_{ONS}$ 
for 2D-SC correlations (blue dashed line in Fig.\,\ref{phdiag008}). Notice that  zero resistance shown by the blue 
region in Fig.\,\ref{phdiag008} occurs at a much lower temperature (red dashed line) corresponding to 3D-phase 
coherence. This issue is discussed below. 

Coming back to $T_{ONS}$, we see that it gets rapidly suppressed as a function of field and points to a 
CQCP around $H_{CQCP}=22$T for sample 008 ($x=0.08$). See Fig.\,\ref{phdiag008} (a) and (b).  
The existence of this CQCP (and an associated plateau) was already proposed, together with the existence 
of a two-stage transition in Ref.\,\cite{Leridon2013}. As also observed in this previous work, the resistivity 
per square at this plateau corresponds quite closely to the quantum of resistance $R_Q$ indicated by an open 
circle at the origin in Figs.\,\ref{008rho}(a) and (b).  $R_{\square}=R_Q$ is also indicated with 
a white dashed line in Fig.\,\ref{phdiag008}. Theory predicts that in the case of a perfectly self-dual 
insulator-SC transition, the critical resistance should be equal to the quantum of resistance \cite{Fisher1990a, Fisher1990b, Cha1991}. Real systems may show deviation from perfect duality and a different critical resistance.
Interestingly enough, the present way of plotting data reveals that 
$R_Q$ indeed coincides with the separatrix line over a broad temperature and field range. A study in which the carrier 
density in a single layer of the same material was tuned with an electric field \cite{Bollinger2011} found 
$R_{\square}=R_Q$ at the insulator-SC transition, in good agreement with the critical resistance in the present 
study. It is worth mentioning that in the latter experiment the lowest temperature measured was 4.3\,K, so the FSC 
observed here was not accessible. 

At high fields and below $T\approx 60$\,K, the polycrystalline CDW phase becomes relevant.
The crossover is characterized by a change from metallic behavior at high temperature to semiconducting behavior 
at low temperature, justifying the choice $T_{min}\sim T_{CDW}$ as the characteristic CDW onset temperature 
(black dashed line). We see that $T_{CDW}$ approximately mirrors the behavior of $T_{ONS}$ and drops dramatically 
with decreasing field in the critical region.  

At low temperatures and intermediate fields a maximum of the resistivity appears preceded by a nearby inflection 
point [see for example the blue curve at intermediate field in Fig.\,\ref{008rho}(a),(b)] and Fig.\,S2. We associate 
this behavior with the onset of FSC transition $T_{ONSF}$. Indeed, $T_{ONSF}$ as function of field shows the expected 
characteristic  ``foot'' in the phase diagram of Fig.\,\ref{phdiag008} [see Figs.\,\ref{fig:phasestif}(a) and 
\ref{fig:tcdg}(b)]. Another fingerprint of FSC is that both $T_{inf}$ (dashed blue line) and $T_{max}$ (dashed black 
line) are very close in this region. 

In order to further check the experimental result we have repeated the analysis on a different LSCO thin film with 
slightly different Sr content  $x=0.09$ but different growing conditions [see Fig.\,S7 in Supplemental Material and 
Fig.\,\ref{phdiag008}(c)].    

Fig.\,\ref{phdiag008}(c) shows the phase diagram for sample 009 with $x=0.09$ and Fig.\,S7 shows the fit to 
the transport data. In this case the drop of $T_{ONS}$ is more gradual, which could be related to a more gradual 
decrease of the stiffness when disorder is increased(compare $T_{ONS}, T_{ONSF}$ with the light blue curve in Fig.\,\ref{fig:phasestif}).  
In any case, since samples 008 and 009 are grown under different conditions and on different substrates (see  
Methods), the fact that the FSC is observed in both cases pleads for the universality of the phase diagram in
this region of doping.

\section*{Discussion}

We propose that in systems in which attractive interactions drive 
CDW and superconductivity with similar energies, disorder can induce FSC when the CDW is more stable in the clean 
limit. Long-range superconductivity eventually takes place when the temperature is low enough to allow the 
phase locking between the SC interfaces. 

We have taken a simplified model based on a CDW with only two possible
variants or ``colors'', A/B. We expect that increasing the number of CDW colors does not change substantially 
our theoretical results.  Also in our modeling we do not need to specify the microscopic origin of CDW colors. There 
are presently several possibilities in the case of cuprates which we now discuss: i) Scanning tunneling 
microscopy\cite{Mesaros2016} have shown that underdoped cuprates are characterized by CDW domains with 4-lattice 
spacing periodicity separated by discommensurations. This naturally produces 4 colors for the CDW for a given 
orientation of the unidirectional CDW  (see Fig.\,3F in Ref.\,\cite{Mesaros2016}). ii) A related possibility is 
that stripes are formed at high temperatures but are metallic and half-filled \cite{Lorenzana2002a} and develop 
a secondary CDW Peierls stability along the stripe which, in strong coupling, can be seen as a lattice of Cooper 
pairs \cite{Bosch2001}. It is natural to describe this state with an effective negative-$U$ Hubbard description 
along chains with an associate quasidegenerate SC state. iii) Yet another possibility is suggested by a microscopic 
analysis which finds an incommensurate CDW in oxygen with $d$-wave symmetry which can be rotated to $d$-wave 
superconductivity \cite{Sachdev2013,Efetov2013}. Here, an Ising order parameter controls excess charge in $x$ 
oriented O bonds with respect to $y$ oriented O bonds, which can be associated to the two possible 
colors of our description. iv) Alternatively, one can see the incommensurate nature as consequence of the weak 
coupling analysis and consider a locally commensurate (strong coupling) version of the theory with 
superconductivity nucleating at the discommensurations, as in i). More experimental and theoretical work is needed 
to establish which scenario occurs in a particular material.

The disorder-induced coexistence of superconductivity and CDW can be seen as a form of intertwined order in the 
sense of Ref.\,\cite{Fradkin2015}. However, these authors treat pair-density-wave order (a self-organized version 
of the Fulde-Ferell-Larkin-Ovchinikov state \cite{Larkin1964,Fulde1964}) as the primary order and CDW as a parasitic 
order. In the present scenario, both CDW and bulk SC order are primary, while FSC is parasitic.    

We have used transport data to derive a phase diagram assuming the magnetic field as tuning parameter 
(Fig.\,\ref{fig:tcdg}, right). We expect similar phase diagrams using lattice strain \cite{Kim2018}, 
field effect \cite{Bollinger2011}, or simply doping as tuning parameters. Indeed, comparing \ref{phdiag008}(a) 
and (c), one concludes that doping plays a similar role to that of magnetic field, since the phase diagram 
appears rigidly shifted. The advantage of the magnetic field is that, being associated with a small energy scale, 
a high resolution scan of the crossovers is possible. 

The main difference between the theoretical and the experimental phase diagram is that superconductivity in the 
former is replaced by 2D fluctuating superconductivity in the latter. One should take into account that the theory 
does not include long-range interactions and quantum fluctuations which are expected to suppress the zero 
resistance state \cite{Emery1995}. Therefore, we  associate $T_{ONS}$ to the transition temperature of the model 
without these effects. With this caveat, the two phase diagrams are in excellent agreement, in particular, the 
experimental phase diagram clearly exhibits the foot-like behavior indicating FSC. 

After the theory part of this work was completed and possted in Ref.~\cite{Attanasi2008},  Ref.\,\cite{yu2019} appeared where 
a very similar phase diagram was derived in a model of superconductivity competing with incommensurate CDW (rather 
as commensurate as here) in the presence of disorder.  The kind of topological defects considered are different - the latter model does not predict FSC. Nevertheless the fact that the essential physical outcomes of the two approaches 
are similar pleads in favour of a rather generic character of disorder-induced SC inside an otherwise stable CDW phase.

It has been proposed that in some underdoped cuprates long-range SC order \cite{Himeda2002,Li2007,Berg2007} is 
frustrated by a peculiar symmetry of the SC state. It is not clear at the moment if this effect contributes also 
to the difference between $T_{ONS}$ and $T_c$ in the present samples. One can reverse the argument and argue that 
FSC is particularly unsuited for 3D phase locking, as the filaments in one plane will in general not coincide with 
the filaments in the next plane, thus frustrating Josephson coupling. 

The deduced phase diagram is not peculiar of the sample analyzed in Fig.\,\ref{phdiag008}. Remarkably, an almost identical phase diagram has 
been derived by completely different techniques in a different material, namely specific heat measurements 
\cite{Kacmarcik2018} in  YBa$_2$Cu$_3$O$_y$ suggesting that this phase diagram is a quite generic feature of 
underdoped cuprates. 

The model presented here is very general and applies to other systems as well, where the balance between CDW and 
SC can produce a topologically protected intertwined order. A particularly interesting model system is 
the Cu-intercalated dichalcogenide 1T-TiSe$_2$.  In this system scanning tunneling microscopy \cite{Yan2017} shows 
a commensurate CDW in the undoped system with domain walls appearing upon Cu intercalation. 
Simultaneously with the latter, superconductivity appears too. The link between CDW discomensurations and 
superconductivity emerges also from magnetoresistance experiments in gated 2D materials \cite{Li2016a,Li2018} and 
from X-ray experiments under pressure \cite{Joe2014}.  
FSC has also been inferred by phenomenological analyses of transport in LaAlO$_3$/SrTiO$_3$ heterostructures\cite{Caprara2013},
 in ZrNCl and some dichalcogenides (TiSe$_2$, MoS$_2$)\cite{Dezi2018}
This adds to the case of $^4$He and suggest that the phenomenon 
at hand is very general. In solids, remarkably, melting occurs first at the surface \cite{Frenken1985}. Thus, when 
a polycrystal is driven just below the melting temperature, the less stable liquid phase nucleates at the interface, 
producing a classical analog of FSC/superfluidity and underlying again the generality of the phenomenon. 

\section*{Methods}

\subsection*{Numerical Simulations}
In order to obtain the stiffness as a function of disorder and tuning parameter (Fig.\,\ref{fig:phasestif}), the 
energy of the model was minimized for configurations in which the random fields were chosen with a flat probability 
distribution between $-W$ and $W$, using a steepest descent algorithm. Then the stiffness was found using linear response and mapping to a resistor network. 
Result for each parameter were averaged over 
200 different configurations. More detail on the computations can be find in Ref.\,\cite{Attanasi2008}. 

\subsection*{Transport experiments}
The thin film for sample 008 was deposited onto STO substrates at KU Leuven, using dc magnetron sputtering 
as described in Ref.\,\cite{LeridonPRB2007}. For this sample, the resistance as function of magnetic field for 
different temperatures was measured in KU Leuven high pulsed magnetic field facilities, using four probe measurements 
on an epitaxial film of thickness $t=100$\,nm, patterned in strips of 1\,mm$\times\,500\,\mu$m. 
High-field pulses up to 49\,T were applied from 1.5 to 300\,K perpendicularly to the ab-plane of the c-axis oriented 
films. We therefore obtained a set of $R_\square(H,T)$ data. The 009 thin film was grown by pulsed laser deposition 
in IIT Kanpur on LSAO substrate. Similar $R_\square(H,T)$ data as a function of field up to 54\,T and temperature 
from 1.5 to 300\,K was obtained at LNCMI Toulouse high field facility.

\subsection*{Fitting procedure}
 
$R_{\square,{\rm CO}}(T)$ in Eq.\,(\ref{eq:rdt}) was obtained by fitting the resistivity at the highest 
field measured (typically $H=50$\,T, black data in Fig.\,\ref{008rho}) and maintained for all the fields measured 
in that sample. For $R_{\square,{\rm CO}}(T)$ typically we used a linear term plus a polynomial in $\log(T)$ up 
to $\log(T)^3$. 

Once $R_{\square,{\rm CO}}(T)$ was fixed, fits were done for the full set of data in the $H,T$ plane 
simultaneously minimizing the total square error. To define the 2D fitting functions, we took the parameters in 
the fit namely $T_c^B$, $T_c^F$,$T_0$, $T_1$ to be polynomials in $H$ of degree 3,1,3,2 respectively. For the 
parameter $\sigma_{\square,0}(H)$ we used a Lorentzian in $H$ centered at $H=0$. Because scans where done in field 
at fixed temperatures, the data used has very high resolution in field (nearly 1300 field values between 0 and 50\,T) 
and much lower resolution in temperature (16 temperatures with higher resolution at low temperatures).\\

\section*{Acknowledgments} J.L is very much indebted with Andrea Cavagna and Carlo Di Castro for important discussions at the early 
stages of this work. J.L. and S.C. thank all the colleagues of the ESPCI in Paris for their warm hospitality 
and for many useful discussion while this work was done.  The work at the KU Leuven has been supported 
by the FWO Programmes and Methusalem Funding by the Flemish Government. Research at IIT Kanpur has been supported 
by the J.C. Bose National Fellowship (R.C.B.). Part of this work has been founded by EuroMagNET II under the EU 
contract number 228043. S.C. and M.G. acknowledge financial support of the University of Rome Sapienza, under the Ateneo 2017 (prot. RM11715C642E8370) and Ateneo 2018 (prot. RM11816431DBA5AF projects. Part of the work was supported through the Chaire Joliot at ESPCI Paris. This work was supported by EU through the COST action CA16218.

\bibliographystyle{naturemag_no_url}

\begin{thebibliography}{10}
\expandafter\ifx\csname url\endcsname\relax
  \def\url#1{\texttt{#1}}\fi
\expandafter\ifx\csname urlprefix\endcsname\relax\def\urlprefix{URL }\fi
\providecommand{\bibinfo}[2]{#2}
\providecommand{\eprint}[2][]{\url{#2}}

\bibitem{Nozieres1985}
\bibinfo{author}{Nozi\`eres, P.} \& \bibinfo{author}{Schmitt-Rink, S.}
\newblock \bibinfo{title}{{Bose condensation in an attractive fermion gas: From
  weak to strong coupling superconductivity}}.
\newblock \emph{\bibinfo{journal}{J. Low Temp. Phys.}}
  \textbf{\bibinfo{volume}{59}}, \bibinfo{pages}{195--211}
  (\bibinfo{year}{1985}).
\newblock \urlprefix\url{http://link.springer.com/10.1007/BF00683774}.

\bibitem{Balibar2008}
\bibinfo{author}{Balibar, S.} \& \bibinfo{author}{Caupin, F.}
\newblock \bibinfo{title}{{Supersolidity and disorder}}.
\newblock \emph{\bibinfo{journal}{J. Phys. Condens. Matter}}
  \textbf{\bibinfo{volume}{20}} (\bibinfo{year}{2008}).

\bibitem{Attanasi2008}
\bibinfo{author}{Attanasi, A.}
\newblock \emph{\bibinfo{title}{{Competition between Superconductivity and
  Charge Density Waves: the Role of Disorder, arXiv:0906.1159}}}.
\newblock Ph.D. thesis, \bibinfo{school}{Sapienza Universit{\`{a}} di Roma}
  (\bibinfo{year}{2008}).
\newblock \urlprefix\url{https://arxiv.org/abs/0906.1159}.

\bibitem{Gerber2015}
\bibinfo{author}{Gerber, S.} \emph{et~al.}
\newblock \bibinfo{title}{{Three-dimensional charge density wave order in YBa 2
  Cu 3 O 6.67 at high magnetic fields}}.
\newblock \emph{\bibinfo{journal}{Science (80-. ).}}
  \textbf{\bibinfo{volume}{350}}, \bibinfo{pages}{949--952}
  (\bibinfo{year}{2015}).
\newblock \urlprefix\url{https://science.sciencemag.org/content/350/6263/949}.

\bibitem{Fradkin2015}
\bibinfo{author}{Fradkin, E.}, \bibinfo{author}{Kivelson, S.~A.} \&
  \bibinfo{author}{Tranquada, J.~M.}
\newblock \bibinfo{title}{{Colloquium: Theory of intertwined orders in high
  temperature superconductors}}.
\newblock \emph{\bibinfo{journal}{Rev. Mod. Phys.}}
  \textbf{\bibinfo{volume}{87}}, \bibinfo{pages}{457--482}
  (\bibinfo{year}{2015}).

\bibitem{Micnas1990}
\bibinfo{author}{Micnas, R.}, \bibinfo{author}{Ranninger, J.} \&
  \bibinfo{author}{Robaszkiewicz, S.}
\newblock \bibinfo{title}{{Superconductivity in narrow-band systems with local
  nonretarded attractive interactions}}.
\newblock \emph{\bibinfo{journal}{Rev. Mod. Phys.}}
  \textbf{\bibinfo{volume}{62}}, \bibinfo{pages}{113--171}
  (\bibinfo{year}{1990}).
\newblock \urlprefix\url{https://link.aps.org/doi/10.1103/RevModPhys.62.113}.

\bibitem{Anderson1958}
\bibinfo{author}{Anderson, P.~W.}
\newblock \bibinfo{title}{{Random-phase approximation in the theory of
  superconductivity}}.
\newblock \emph{\bibinfo{journal}{Phys. Rev.}} \textbf{\bibinfo{volume}{112}},
  \bibinfo{pages}{1900--1916} (\bibinfo{year}{1958}).
\newblock \urlprefix\url{http://dx.doi.org/10.1103/physrev.112.1900}.

\bibitem{Liu1972}
\bibinfo{author}{Liu, K.-S.} \& \bibinfo{author}{Fisher, M.~E.}
\newblock \bibinfo{title}{{Quantum Lattice Gas and the Existence of a
  Supersolid}}.
\newblock \emph{\bibinfo{journal}{J. Low Temp. Phys.}}
  \textbf{\bibinfo{volume}{10}}, \bibinfo{pages}{655} (\bibinfo{year}{1972}).

\bibitem{Cha1991}
\bibinfo{author}{Cha, M.~C.}, \bibinfo{author}{Fisher, M.~P.},
  \bibinfo{author}{Girvin, S.~M.}, \bibinfo{author}{Wallin, M.} \&
  \bibinfo{author}{Young, A.~P.}
\newblock \bibinfo{title}{{Universal conductivity of two-dimensional films at
  the superconductor- insulator transition}}.
\newblock \emph{\bibinfo{journal}{Phys. Rev. B}} \textbf{\bibinfo{volume}{44}},
  \bibinfo{pages}{6883--6902} (\bibinfo{year}{1991}).

\bibitem{Binder1983}
\bibinfo{author}{Binder, K.}
\newblock \bibinfo{title}{{Random-field induced interface widths in Ising
  systems}}.
\newblock \emph{\bibinfo{journal}{Zeitschrift f{\"{u}}r Phys. B Condens.
  Matter}} \textbf{\bibinfo{volume}{50}}, \bibinfo{pages}{343--352}
  (\bibinfo{year}{1983}).

\bibitem{Imry1975}
\bibinfo{author}{Imry, Y.} \& \bibinfo{author}{Ma, S.-k.}
\newblock \bibinfo{title}{{Random-Field Instability of the Ordered State of
  Continuous Symmetry}}.
\newblock \emph{\bibinfo{journal}{Phys. Rev. Lett.}}
  \textbf{\bibinfo{volume}{35}}, \bibinfo{pages}{1399--1401}
  (\bibinfo{year}{1975}).
\newblock \urlprefix\url{http://dx.doi.org/10.1103/PhysRevLett.35.1399
  https://link.aps.org/doi/10.1103/PhysRevLett.35.1399}.

\bibitem{Cuccoli2003a}
\bibinfo{author}{Cuccoli, A.}, \bibinfo{author}{Roscilde, T.},
  \bibinfo{author}{Tognetti, V.}, \bibinfo{author}{Vaia, R.} \&
  \bibinfo{author}{Verrucchi, P.}
\newblock \bibinfo{title}{{Quantum Monte Carlo study of S=1/2 weakly
  anisotropic antiferromagnets on the square lattice}}.
\newblock \emph{\bibinfo{journal}{Phys. Rev. B}} \textbf{\bibinfo{volume}{67}},
  \bibinfo{pages}{104414} (\bibinfo{year}{2003}).
\newblock
  \urlprefix\url{http://link.aps.org/doi/10.1103/PhysRevB.67.104414%5Cnhttps://link.aps.org/doi/10.1103/PhysRevB.67.104414}.
\newblock \eprint{0209316}.

\bibitem{Miao2017}
\bibinfo{author}{Miao, H.} \emph{et~al.}
\newblock \bibinfo{title}{{High-temperature charge density wave correlations in
  La 1.875 Ba 0.125 CuO 4 without spin–charge locking}}.
\newblock \emph{\bibinfo{journal}{Proc. Natl. Acad. Sci.}}
  \textbf{\bibinfo{volume}{114}}, \bibinfo{pages}{12430--12435}
  (\bibinfo{year}{2017}).
\newblock
  \urlprefix\url{http://www.pnas.org/lookup/doi/10.1073/pnas.1708549114}.

\bibitem{Nie2014}
\bibinfo{author}{Nie, L.}, \bibinfo{author}{Tarjus, G.} \&
  \bibinfo{author}{Kivelson, S.~A.}
\newblock \bibinfo{title}{{Quenched disorder and vestigial nematicity in the
  pseudogap regime of the cuprates}}.
\newblock \emph{\bibinfo{journal}{Proc. Natl. Acad. Sci.}}
  \textbf{\bibinfo{volume}{111}}, \bibinfo{pages}{7980--7985}
  (\bibinfo{year}{2014}).
\newblock \urlprefix\url{http://arxiv.org/abs/1311.5580
  http://arxiv.org/abs/1311.5580%0Ahttp://dx.doi.org/10.1073/pnas.1406019111
  http://www.pnas.org/cgi/doi/10.1073/pnas.1406019111
  http://www.pubmedcentral.nih.gov/articlerender.fcgi?artid=4050631&tool=pmcentrez&rendertype=abstr}.
\newblock \eprint{1311.5580}.

\bibitem{Capati2015}
\bibinfo{author}{Capati, M.} \emph{et~al.}
\newblock \bibinfo{title}{{Electronic polymers and soft-matter-like broken
  symmetries in underdoped cuprates}}.
\newblock \emph{\bibinfo{journal}{Nat. Commun.}} \textbf{\bibinfo{volume}{6}},
  \bibinfo{pages}{10} (\bibinfo{year}{2015}).
\newblock \urlprefix\url{http://www.nature.com/articles/ncomms8691
  http://arxiv.org/abs/1505.01847}.
\newblock \eprint{1505.01847}.

\bibitem{Ando2004}
\bibinfo{author}{Ando, Y.}, \bibinfo{author}{Komiya, S.},
  \bibinfo{author}{Segawa, K.}, \bibinfo{author}{Ono, S.} \&
  \bibinfo{author}{Kurita, Y.}
\newblock \bibinfo{title}{{Electronic Phase Diagram of High-$T_c$ Cuprate
  Superconductors from a Mapping of the In-Plane Resistivity Curvature}}.
\newblock \emph{\bibinfo{journal}{Phys. Rev. Lett.}}
  \textbf{\bibinfo{volume}{93}}, \bibinfo{pages}{267001}
  (\bibinfo{year}{2004}).
\newblock
  \urlprefix\url{http://arxiv.org/abs/cond-mat/0403032%0Ahttp://dx.doi.org/10.1103/PhysRevLett.93.267001
  https://link.aps.org/doi/10.1103/PhysRevLett.93.267001}.
\newblock \eprint{0403032}.

\bibitem{Pelc2017}
\bibinfo{author}{Pelc, D.}, \bibinfo{author}{Pop{\v{c}}evi{\'{c}}, P.},
  \bibinfo{author}{Po{\v{z}}ek, M.}, \bibinfo{author}{Greven, M.} \&
  \bibinfo{author}{Bari{\v{s}}i{\'{c}}, N.}
\newblock \bibinfo{title}{{Unusual behavior of cuprates explained by
  heterogeneous charge localization}}.
\newblock \emph{\bibinfo{journal}{Sci. Adv.}} \textbf{\bibinfo{volume}{5}},
  \bibinfo{pages}{eaau4538} (\bibinfo{year}{2019}).
\newblock \urlprefix\url{http://arxiv.org/abs/1710.10221
  http://advances.sciencemag.org/lookup/doi/10.1126/sciadv.aau4538}.
\newblock \eprint{1710.10221}.

\bibitem{Kim2018}
\bibinfo{author}{Kim, H.-H.} \emph{et~al.}
\newblock \bibinfo{title}{{Uniaxial pressure control of competing orders in a
  high-temperature superconductor}}.
\newblock \emph{\bibinfo{journal}{Science (80-. ).}}
  \textbf{\bibinfo{volume}{362}}, \bibinfo{pages}{1040--1044}
  (\bibinfo{year}{2018}).
\newblock \urlprefix\url{http://www.ncbi.nlm.nih.gov/pubmed/30498124}.

\bibitem{Kivelson2003}
\bibinfo{author}{Kivelson, S.~A.} \emph{et~al.}
\newblock \bibinfo{title}{{How to detect fluctuating stripes in the
  high-temperature superconductors}}.
\newblock \emph{\bibinfo{journal}{Rev. Mod. Phys.}}
  \textbf{\bibinfo{volume}{75}}, \bibinfo{pages}{1201--1241}
  (\bibinfo{year}{2003}).

\bibitem{Wu2011}
\bibinfo{author}{Wu, T.} \emph{et~al.}
\newblock \bibinfo{title}{{Magnetic-field-induced charge-stripe order in the
  high-temperature superconductor YBa2Cu3O y}}.
\newblock \emph{\bibinfo{journal}{Nature}} \textbf{\bibinfo{volume}{477}},
  \bibinfo{pages}{191--194} (\bibinfo{year}{2011}).
\newblock \urlprefix\url{http://www.nature.com/articles/nature10345}.

\bibitem{Doiron-Leyraud2007}
\bibinfo{author}{Doiron-Leyraud, N.} \emph{et~al.}
\newblock \bibinfo{title}{{Quantum oscillations and the Fermi surface in an
  underdoped high-Tc superconductor.}}
\newblock \emph{\bibinfo{journal}{Nature}} \textbf{\bibinfo{volume}{447}},
  \bibinfo{pages}{565--8} (\bibinfo{year}{2007}).
\newblock \urlprefix\url{http://www.ncbi.nlm.nih.gov/pubmed/17538614}.

\bibitem{Aslamazov1968a}
\bibinfo{author}{Aslamazov, L.~G.} \& \bibinfo{author}{Larkin, A.~I.}
\newblock \bibinfo{title}{{Effect of fluctuations on the properties of a
  superconductor above the critical temperature}}.
\newblock \emph{\bibinfo{journal}{Sov. Phys. - Solid State}}
  \textbf{\bibinfo{volume}{10}}, \bibinfo{pages}{875} (\bibinfo{year}{1968}).

\bibitem{Aslamasov1968}
\bibinfo{author}{Aslamasov, L.~G.} \& \bibinfo{author}{Larkin, A.~I.}
\newblock \bibinfo{title}{{The influence of fluctuation pairing of electrons on
  the conductivity of normal metal}}.
\newblock \emph{\bibinfo{journal}{Phys. Lett. A}}
  \textbf{\bibinfo{volume}{26}}, \bibinfo{pages}{238--239}
  (\bibinfo{year}{1968}).
\newblock
  \urlprefix\url{https://www.sciencedirect.com/science/article/abs/pii/0375960168906233}.

\bibitem{LeridonPRB2007}
\bibinfo{author}{Leridon, B.}, \bibinfo{author}{Vanacken, J.},
  \bibinfo{author}{Wambecq, T.} \& \bibinfo{author}{Moshchalkov, V.~V.}
\newblock \bibinfo{title}{Paraconductivity of underdoped
  $\mbox{La}_{2-x}\mbox{Sr}_x\mbox{CuO}_4$ thin-film superconductors using high
  magnetic fields}.
\newblock \emph{\bibinfo{journal}{Physical Review B}}
  \textbf{\bibinfo{volume}{76}} (\bibinfo{year}{2007}).
\newblock \urlprefix\url{https://link.aps.org/doi/10.1103/PhysRevB.76.012503}.

\bibitem{Caprara2009}
\bibinfo{author}{Caprara, S.}, \bibinfo{author}{Grilli, M.},
  \bibinfo{author}{Leridon, B.} \& \bibinfo{author}{Vanacken, J.}
\newblock \bibinfo{title}{{Paraconductivity in layered cuprates behaves as if
  due to pairing of nearly free quasiparticles}}.
\newblock \emph{\bibinfo{journal}{Phys. Rev. B - Condens. Matter Mater. Phys.}}
  \textbf{\bibinfo{volume}{79}}, \bibinfo{pages}{024506}
  (\bibinfo{year}{2009}).
\newblock \urlprefix\url{https://link.aps.org/doi/10.1103/PhysRevB.79.024506}.

\bibitem{LeridonPRL2001}
\bibinfo{author}{Leridon, B.}, \bibinfo{author}{Defossez, A.},
  \bibinfo{author}{Dumont, J.}, \bibinfo{author}{Lesueur, J.} \&
  \bibinfo{author}{Contour, J.~P.}
\newblock \bibinfo{title}{Conductivity of underdoped
  $\mbox{YBa}_2\mbox{Cu}_3\mbox{O}_{7 -?}$ : Evidence for incoherent pair
  correlations in the pseudogap regime}.
\newblock \emph{\bibinfo{journal}{Physical Review Letters}}
  \textbf{\bibinfo{volume}{87}} (\bibinfo{year}{2001}).
\newblock
  \urlprefix\url{https://link.aps.org/doi/10.1103/PhysRevLett.87.197007}.

\bibitem{LeridonPRL2003}
\bibinfo{author}{Leridon, B.}, \bibinfo{author}{Defossez, A.},
  \bibinfo{author}{Dumont, J.}, \bibinfo{author}{Lesueur, J.} \&
  \bibinfo{author}{Contour, J.~P.}
\newblock \bibinfo{title}{Leridon \textit{et al.} {Reply}:}.
\newblock \emph{\bibinfo{journal}{Physical Review Letters}}
  \textbf{\bibinfo{volume}{90}} (\bibinfo{year}{2003}).
\newblock
  \urlprefix\url{https://link.aps.org/doi/10.1103/PhysRevLett.90.179704}.

\bibitem{Vidal2002}
\bibinfo{author}{Vidal, F.} \emph{et~al.}
\newblock \bibinfo{title}{{On the consequences of the uncertainty principle on
  the superconducting fluctuations well inside the normal state}}.
\newblock \emph{\bibinfo{journal}{Europhys. Lett.}}
  \textbf{\bibinfo{volume}{59}}, \bibinfo{pages}{754--760}
  (\bibinfo{year}{2002}).
\newblock
  \urlprefix\url{http://stacks.iop.org/0295-5075/59/i=5/a=754?key=crossref.d69a9589872140bb85fb4075a1dfad09}.

\bibitem{Mishonov2003}
\bibinfo{author}{Mishonov, T.~M.}, \bibinfo{author}{Pachov, G.~V.},
  \bibinfo{author}{Genchev, I.~N.}, \bibinfo{author}{Atanasova, L.~A.} \&
  \bibinfo{author}{Damianov, D.~C.}
\newblock \bibinfo{title}{{Kinetics and Boltzmann kinetic equation for
  fluctuation Cooper pairs}}.
\newblock \emph{\bibinfo{journal}{Phys. Rev. B - Condens. Matter Mater. Phys.}}
  \textbf{\bibinfo{volume}{68}}, \bibinfo{pages}{545251--545258}
  (\bibinfo{year}{2003}).

\bibitem{Caprara2005}
\bibinfo{author}{Caprara, S.}, \bibinfo{author}{Grilli, M.},
  \bibinfo{author}{Leridon, B.} \& \bibinfo{author}{Lesueur, J.}
\newblock \bibinfo{title}{Extended paraconductivity regime in underdoped
  cuprates}.
\newblock \emph{\bibinfo{journal}{Physical Review B}}
  \textbf{\bibinfo{volume}{72}} (\bibinfo{year}{2005}).
\newblock \urlprefix\url{https://link.aps.org/doi/10.1103/PhysRevB.72.104509}.

\bibitem{Luo2003}
\bibinfo{author}{Luo, C.~W.} \emph{et~al.}
\newblock \bibinfo{title}{Comment on {Conductivity} of {Underdoped}
  YBa$_2$Cu$_3$O$_{7 -\delta}$ : {Evidence} for {Incoherent} {Pair} {Correlations}
  in the {Pseudogap} {Regime}}
\newblock \emph{\bibinfo{journal}{Physical Review Letters}}
  \textbf{\bibinfo{volume}{90}} (\bibinfo{year}{2003}).
\newblock
  \urlprefix\url{https://link.aps.org/doi/10.1103/PhysRevLett.90.179703}.

\bibitem{popcevic2018}
\bibinfo{author}{Pop\v{c}evi\'{c}, P.} \emph{et~al.}
\newblock \bibinfo{title}{Percolative nature of the direct-current
  paraconductivity in cuprate superconductors}.
\newblock \emph{\bibinfo{journal}{npj Quantum Materials}}
  \textbf{\bibinfo{volume}{3}} (\bibinfo{year}{2018}).
\newblock \urlprefix\url{http://www.nature.com/articles/s41535-018-0115-2}.

\bibitem{Leridon2013}
\bibinfo{author}{Leridon, B.} \emph{et~al.}
\newblock \bibinfo{title}{Double criticality in the magnetic field-driven
  transition of a high-tc superconductor}.
\newblock \emph{\bibinfo{journal}{arXiv:1306.4583}}  (\bibinfo{year}{2013}).
\newblock \urlprefix\url{https://arxiv.org/abs/1306.4583}.

\bibitem{Fisher1990a}
\bibinfo{author}{Fisher, M. P.~A.}, \bibinfo{author}{Grinstein, G.} \&
  \bibinfo{author}{Girvin, S.~M.}
\newblock \bibinfo{title}{Presence of quantum diffusion in two dimensions:
  {Universal} resistance at the superconductor/insulator transition}.
\newblock \emph{\bibinfo{journal}{Physical Review Letters}}
  \textbf{\bibinfo{volume}{64}} (\bibinfo{year}{1990}).

\bibitem{Fisher1990b}
\bibinfo{author}{Fisher, M. P.~A.}
\newblock \bibinfo{title}{Quantum phase transitions in disordered
  two-dimensional superconductors}.
\newblock \emph{\bibinfo{journal}{Physical Review Letters}}
  \textbf{\bibinfo{volume}{65}} (\bibinfo{year}{1990}).

\bibitem{Bollinger2011}
\bibinfo{author}{Bollinger, A.~T.} \emph{et~al.}
\newblock \bibinfo{title}{{Superconductor-insulator transition in
  La2-xSrxCuO4at the pair quantum resistance}}.
\newblock \emph{\bibinfo{journal}{Nature}} \textbf{\bibinfo{volume}{472}},
  \bibinfo{pages}{458--460} (\bibinfo{year}{2011}).

\bibitem{Mesaros2016}
\bibinfo{author}{Mesaros, A.} \emph{et~al.}
\newblock \bibinfo{title}{{Commensurate 4a0-period charge density modulations
  throughout the Bi2Sr2CaCu2O8+x pseudogap regime.}}
\newblock \emph{\bibinfo{journal}{Proc. Natl. Acad. Sci. U. S. A.}}
  \textbf{\bibinfo{volume}{113}}, \bibinfo{pages}{12661--12666}
  (\bibinfo{year}{2016}).
\newblock \urlprefix\url{http://www.ncbi.nlm.nih.gov/pubmed/27791157
  http://www.pubmedcentral.nih.gov/articlerender.fcgi?artid=PMC5111700}.

\bibitem{Lorenzana2002a}
\bibinfo{author}{Lorenzana, J.} \& \bibinfo{author}{Seibold, G.}
\newblock \bibinfo{title}{{Metallic Mean-Field Stripes, Incommensurability, and
  ChemicalPotential in Cuprates}}.
\newblock \emph{\bibinfo{journal}{Phys. Rev. Lett.}}
  \textbf{\bibinfo{volume}{89}}, \bibinfo{pages}{136401}
  (\bibinfo{year}{2002}).
\newblock
  \urlprefix\url{http://link.aps.org/doi/10.1103/PhysRevLett.89.136401}.

\bibitem{Bosch2001}
\bibinfo{author}{Bosch, M.}, \bibinfo{author}{van Saarloos, W.} \&
  \bibinfo{author}{Zaanen, J.}
\newblock \bibinfo{title}{{Shifting Bragg peaks of cuprate stripes as possible
  indications for fractionally charged kinks}}.
\newblock \emph{\bibinfo{journal}{Phys. Rev. B}} \textbf{\bibinfo{volume}{63}},
  \bibinfo{pages}{1--4} (\bibinfo{year}{2001}).
\newblock \urlprefix\url{http://link.aps.org/doi/10.1103/PhysRevB.63.092501}.

\bibitem{Sachdev2013}
\bibinfo{author}{Sachdev, S.} \& \bibinfo{author}{{La Placa}, R.}
\newblock \bibinfo{title}{{Bond order in two-dimensional metals with
  antiferromagnetic exchange interactions}}.
\newblock \emph{\bibinfo{journal}{Phys. Rev. Lett.}}
  \textbf{\bibinfo{volume}{111}}, \bibinfo{pages}{1--5} (\bibinfo{year}{2013}).

\bibitem{Efetov2013}
\bibinfo{author}{Efetov, K.~B.}, \bibinfo{author}{Meier, H.} \&
  \bibinfo{author}{P{\'{e}}pin, C.}
\newblock \bibinfo{title}{{Pseudogap state near a quantum critical point}}.
\newblock \emph{\bibinfo{journal}{Nat. Phys.}} \textbf{\bibinfo{volume}{9}},
  \bibinfo{pages}{442--446} (\bibinfo{year}{2013}).
\newblock \urlprefix\url{http://dx.doi.org/10.1038/nphys2641}.

\bibitem{Larkin1964}
\bibinfo{author}{Larkin, A.~I.} \& \bibinfo{author}{Ovchinnikov, Y.~N.}
\newblock \bibinfo{title}{{Nonuniform state of superconductors}}.
\newblock \emph{\bibinfo{journal}{Zh. Eksperim. i Teor. Fiz.}}
  \textbf{\bibinfo{volume}{47}}, \bibinfo{pages}{1136 [Sov. Phys. JETP. 20, 762
  (1965)]} (\bibinfo{year}{1964}).

\bibitem{Fulde1964}
\bibinfo{author}{Fulde, P.} \& \bibinfo{author}{Ferrell, R.~A.}
\newblock \bibinfo{title}{{Superconductivity in a Strong Spin-Exchange Field}}.
\newblock \emph{\bibinfo{journal}{Phys. Rev.}} \textbf{\bibinfo{volume}{135}},
  \bibinfo{pages}{A550--A563} (\bibinfo{year}{1964}).
\newblock \urlprefix\url{https://link.aps.org/doi/10.1103/PhysRev.135.A550}.

\bibitem{Emery1995}
\bibinfo{author}{Emery, V.~J.} \& \bibinfo{author}{Kivelson, S.~A.}
\newblock \bibinfo{title}{{Importance of phase fluctuations in superconductors
  with small superfluid density}}.
\newblock \emph{\bibinfo{journal}{Nature}} \textbf{\bibinfo{volume}{374}},
  \bibinfo{pages}{434--437} (\bibinfo{year}{1995}).
\newblock \urlprefix\url{http://www.nature.com/articles/374434a0}.

\bibitem{yu2019}
\bibinfo{author}{Yu, Y.} \& \bibinfo{author}{Kivelson, S.~A.}
\newblock \bibinfo{title}{{Fragile superconductivity in the presence of weakly
  disordered charge density waves}}.
\newblock \emph{\bibinfo{journal}{Phys. Rev. B}} \textbf{\bibinfo{volume}{99}},
  \bibinfo{pages}{144513} (\bibinfo{year}{2019}).
\newblock \urlprefix\url{https://link.aps.org/doi/10.1103/PhysRevB.99.144513}.

\bibitem{Himeda2002}
\bibinfo{author}{Himeda, A.}, \bibinfo{author}{Kato, T.} \&
  \bibinfo{author}{Ogata, M.}
\newblock \bibinfo{title}{{Stripe States with Spatially Oscillating [Formula
  presented]-Wave Superconductivity in the Two-Dimensional [Formula presented]
  Model}}.
\newblock \emph{\bibinfo{journal}{Phys. Rev. Lett.}}
  \textbf{\bibinfo{volume}{88}}, \bibinfo{pages}{4} (\bibinfo{year}{2002}).

\bibitem{Li2007}
\bibinfo{author}{Li, Q.}, \bibinfo{author}{H{\"{u}}cker, M.},
  \bibinfo{author}{Gu, G.~D.}, \bibinfo{author}{Tsvelik, A.~M.} \&
  \bibinfo{author}{Tranquada, J.~M.}
\newblock \bibinfo{title}{{Two-Dimensional Superconducting Fluctuations in
  Stripe-Ordered La$_{1.875}$Ba$_{0.125}$CuO$_4$}}.
\newblock \emph{\bibinfo{journal}{Phys. Rev. Lett.}}
  \textbf{\bibinfo{volume}{99}}, \bibinfo{pages}{067001qq}
  (\bibinfo{year}{2007}).
\newblock
  \urlprefix\url{https://link.aps.org/doi/10.1103/PhysRevLett.99.067001}.

\bibitem{Berg2007}
\bibinfo{author}{Berg, E.} \emph{et~al.}
\newblock \bibinfo{title}{{Dynamical Layer Decoupling in a Stripe-Ordered
  High-T$_c$ Superconductor}}.
\newblock \emph{\bibinfo{journal}{Phys. Rev. Lett.}}
  \textbf{\bibinfo{volume}{99}}, \bibinfo{pages}{127003}
  (\bibinfo{year}{2007}).
\newblock
  \urlprefix\url{https://link.aps.org/doi/10.1103/PhysRevLett.99.127003}.

\bibitem{Kacmarcik2018}
\bibinfo{author}{Ka{\v{c}}mar{\v{c}}{\'{i}}k, J.} \emph{et~al.}
\newblock \bibinfo{title}{{Unusual Interplay between Superconductivity and
  Field-Induced Charge Order in YBa2Cu3 Oy}}.
\newblock \emph{\bibinfo{journal}{Phys. Rev. Lett.}}
  \textbf{\bibinfo{volume}{121}}, \bibinfo{pages}{1--6} (\bibinfo{year}{2018}).

\bibitem{Yan2017}
\bibinfo{author}{Yan, S.} \emph{et~al.}
\newblock \bibinfo{title}{{Influence of Domain Walls in the Incommensurate
  Charge Density Wave State of Cu Intercalated 1T-TiSe$_2$}}.
\newblock \emph{\bibinfo{journal}{Phys. Rev. Lett.}}
  \textbf{\bibinfo{volume}{118}}, \bibinfo{pages}{106405}
  (\bibinfo{year}{2017}).
\newblock
  \urlprefix\url{https://link.aps.org/doi/10.1103/PhysRevLett.118.106405}.

\bibitem{Li2016a}
\bibinfo{author}{Li, L.~J.} \emph{et~al.}
\newblock \bibinfo{title}{{Controlling many-body states by the electric-field
  effect in a two-dimensional material}}.
\newblock \emph{\bibinfo{journal}{Nature}} \textbf{\bibinfo{volume}{529}},
  \bibinfo{pages}{185--189} (\bibinfo{year}{2016}).

\bibitem{Li2018}
\bibinfo{author}{Li, L.} \emph{et~al.}
\newblock \bibinfo{title}{{Anomalous quantum metal in a 2D crystalline
  superconductor with intrinsic electronic non-uniformity}}
  (\bibinfo{year}{2018}).
\newblock \urlprefix\url{http://arxiv.org/abs/1803.10936}.
\newblock \eprint{1803.10936}.

\bibitem{Joe2014}
\bibinfo{author}{Joe, Y.~I.} \emph{et~al.}
\newblock \bibinfo{title}{{Emergence of charge density wave domain walls above
  the superconducting dome in 1T-TiSe 2}}.
\newblock \emph{\bibinfo{journal}{Nat. Phys.}} \textbf{\bibinfo{volume}{10}},
  \bibinfo{pages}{421--425} (\bibinfo{year}{2014}).
\newblock \urlprefix\url{http://www.nature.com/doifinder/10.1038/nphys2935}.
\newblock \eprint{1309.4051}.

\bibitem{Caprara2013}
\bibinfo{author}{Caprara, S.} \emph{et~al.}
\newblock \bibinfo{title}{Multiband superconductivity and nanoscale
  inhomogeneity at oxide interfaces}.
\newblock \emph{\bibinfo{journal}{Physical Review B}}
  \textbf{\bibinfo{volume}{88}} (\bibinfo{year}{2013}).

\bibitem{Dezi2018}
\bibinfo{author}{Dezi, G.}, \bibinfo{author}{Scopigno, N.},
  \bibinfo{author}{Caprara, S.} \& \bibinfo{author}{Grilli, M.}
\newblock \bibinfo{title}{Negative electronic compressibility and nanoscale
  inhomogeneity in ionic-liquid gated two-dimensional superconductors}.
\newblock \emph{\bibinfo{journal}{Physical Review B}}
  \textbf{\bibinfo{volume}{98}} (\bibinfo{year}{2018}).

\bibitem{Frenken1985}
\bibinfo{author}{Frenken, J. W.~M.} \& \bibinfo{author}{{Van Der Veen}, J.~F.}
\newblock \bibinfo{title}{{Observation of surface melting}}.
\newblock \emph{\bibinfo{journal}{Phys. Rev. Lett.}}
  \textbf{\bibinfo{volume}{54}}, \bibinfo{pages}{134--137}
  (\bibinfo{year}{1985}).

\end{thebibliography}

\newpage

\section*{Supplementary Information}
\subsection*{Supplementary Figure}

\begin{figure}[h]
\includegraphics*[width=8cm, clip]{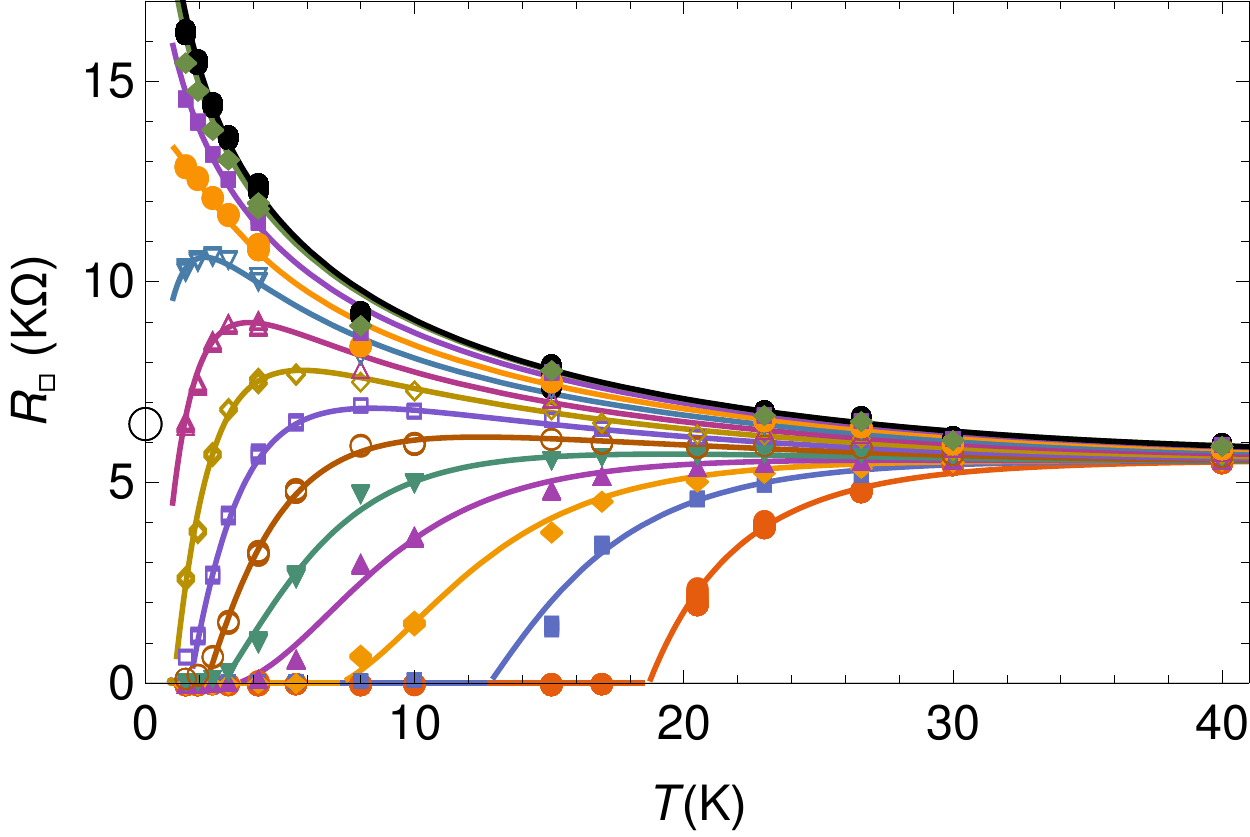}
\caption{Temperature dependent resistance per square for sample 009 (LSCO/LSAO,
$x=0.09$) for different fields. Fields start from $H=0$\,T in the lower curve up to 48\,T in intervals of 4\,T.
The black curve is the fit of the data without SC component at 54\,T (black dots). The occurrence of FSC is quite 
apparent from the low temperature drop of resistance in the curves showing semiconducting behavior.  
The open black circle at $T=0$ indicates the quantum of resistance $R_Q=h/(2e)^2$. The corresponding phase diagram 
is shown in Fig.\,6(c) of the main manuscript. 
} 

\label{009brho}
\end{figure}

\end{document}